\documentclass[twocolumn,
               trackchanges]{aastex631}
\usepackage[utf8]{inputenc}
\usepackage[T5,T1]{fontenc}
\usepackage{savesym}
\usepackage{amsmath}
\usepackage{booktabs}
\usepackage{longtable}
\usepackage{url}

\newcommand{\gcnlink}[1]{\href{https://gcn.nasa.gov/circulars/#1}{{#1}}}
\newcommand{\eiso}{$E_{\mathrm{iso}}$}

\begin{document}

\title{IceCube Real-time Searches for High-energy Neutrinos Coincident with LIGO/Virgo/KAGRA Gravitational-Wave Alerts in O4a}
\correspondingauthor{The IceCube Collaboration}
\email{analysis@icecube.wisc.edu}

\affiliation{III. Physikalisches Institut, RWTH Aachen University, D-52056 Aachen, Germany}
\affiliation{Department of Physics, University of Adelaide, Adelaide, 5005, Australia}
\affiliation{Dept. of Physics and Astronomy, University of Alaska Anchorage, 3211 Providence Dr., Anchorage, AK 99508, USA}
\affiliation{School of Physics and Center for Relativistic Astrophysics, Georgia Institute of Technology, Atlanta, GA 30332, USA}
\affiliation{Dept. of Physics, Southern University, Baton Rouge, LA 70813, USA}
\affiliation{Dept. of Physics, University of California, Berkeley, CA 94720, USA}
\affiliation{Lawrence Berkeley National Laboratory, Berkeley, CA 94720, USA}
\affiliation{Institut f{\"u}r Physik, Humboldt-Universit{\"a}t zu Berlin, D-12489 Berlin, Germany}
\affiliation{Fakult{\"a}t f{\"u}r Physik {\&} Astronomie, Ruhr-Universit{\"a}t Bochum, D-44780 Bochum, Germany}
\affiliation{Universit{\'e} Libre de Bruxelles, Science Faculty CP230, B-1050 Brussels, Belgium}
\affiliation{Vrije Universiteit Brussel (VUB), Dienst ELEM, B-1050 Brussels, Belgium}
\affiliation{Dept. of Physics, Simon Fraser University, Burnaby, BC V5A 1S6, Canada}
\affiliation{Department of Physics and Laboratory for Particle Physics and Cosmology, Harvard University, Cambridge, MA 02138, USA}
\affiliation{Dept. of Physics, Massachusetts Institute of Technology, Cambridge, MA 02139, USA}
\affiliation{Dept. of Physics and The International Center for Hadron Astrophysics, Chiba University, Chiba 263-8522, Japan}
\affiliation{Department of Physics, Loyola University Chicago, Chicago, IL 60660, USA}
\affiliation{Dept. of Physics and Astronomy, University of Canterbury, Private Bag 4800, Christchurch, New Zealand}
\affiliation{Dept. of Physics, University of Maryland, College Park, MD 20742, USA}
\affiliation{Dept. of Astronomy, Ohio State University, Columbus, OH 43210, USA}
\affiliation{Dept. of Physics and Center for Cosmology and Astro-Particle Physics, Ohio State University, Columbus, OH 43210, USA}
\affiliation{Niels Bohr Institute, University of Copenhagen, DK-2100 Copenhagen, Denmark}
\affiliation{Dept. of Physics, TU Dortmund University, D-44221 Dortmund, Germany}
\affiliation{Dept. of Physics and Astronomy, Michigan State University, East Lansing, MI 48824, USA}
\affiliation{Dept. of Physics, University of Alberta, Edmonton, Alberta, T6G 2E1, Canada}
\affiliation{Erlangen Centre for Astroparticle Physics, Friedrich-Alexander-Universit{\"a}t Erlangen-N{\"u}rnberg, D-91058 Erlangen, Germany}
\affiliation{Physik-department, Technische Universit{\"a}t M{\"u}nchen, D-85748 Garching, Germany}
\affiliation{D{\'e}partement de physique nucl{\'e}aire et corpusculaire, Universit{\'e} de Gen{\`e}ve, CH-1211 Gen{\`e}ve, Switzerland}
\affiliation{Dept. of Physics and Astronomy, University of Gent, B-9000 Gent, Belgium}
\affiliation{Dept. of Physics and Astronomy, University of California, Irvine, CA 92697, USA}
\affiliation{Karlsruhe Institute of Technology, Institute for Astroparticle Physics, D-76021 Karlsruhe, Germany}
\affiliation{Karlsruhe Institute of Technology, Institute of Experimental Particle Physics, D-76021 Karlsruhe, Germany}
\affiliation{Dept. of Physics, Engineering Physics, and Astronomy, Queen's University, Kingston, ON K7L 3N6, Canada}
\affiliation{Department of Physics {\&} Astronomy, University of Nevada, Las Vegas, NV 89154, USA}
\affiliation{Nevada Center for Astrophysics, University of Nevada, Las Vegas, NV 89154, USA}
\affiliation{Dept. of Physics and Astronomy, University of Kansas, Lawrence, KS 66045, USA}
\affiliation{UCLouvain, Centre for Cosmology, Particle Physics and Phenomenology, CP3, Chemin du Cyclotron 2, 1348 Louvain-la-Neuve, Belgium}
\affiliation{Department of Physics, Mercer University, Macon, GA 31207-0001, USA}
\affiliation{Dept. of Astronomy, University of Wisconsin{\textemdash}Madison, Madison, WI 53706, USA}
\affiliation{Dept. of Physics and Wisconsin IceCube Particle Astrophysics Center, University of Wisconsin{\textemdash}Madison, Madison, WI 53706, USA}
\affiliation{Institute of Physics, University of Mainz, Staudinger Weg 7, D-55099 Mainz, Germany}
\affiliation{Department of Physics, Marquette University, Milwaukee, WI 53201, USA}
\affiliation{Institut f{\"u}r Kernphysik, Universit{\"a}t M{\"u}nster, D-48149 M{\"u}nster, Germany}
\affiliation{Bartol Research Institute and Dept. of Physics and Astronomy, University of Delaware, Newark, DE 19716, USA}
\affiliation{Dept. of Physics, Yale University, New Haven, CT 06520, USA}
\affiliation{Columbia Astrophysics and Nevis Laboratories, Columbia University, New York, NY 10027, USA}
\affiliation{Dept. of Physics, University of Oxford, Parks Road, Oxford OX1 3PU, United Kingdom}
\affiliation{Dipartimento di Fisica e Astronomia Galileo Galilei, Universit{\`a} Degli Studi di Padova, I-35122 Padova PD, Italy}
\affiliation{Dept. of Physics, Drexel University, 3141 Chestnut Street, Philadelphia, PA 19104, USA}
\affiliation{Physics Department, South Dakota School of Mines and Technology, Rapid City, SD 57701, USA}
\affiliation{Dept. of Physics, University of Wisconsin, River Falls, WI 54022, USA}
\affiliation{Dept. of Physics and Astronomy, University of Rochester, Rochester, NY 14627, USA}
\affiliation{Department of Physics and Astronomy, University of Utah, Salt Lake City, UT 84112, USA}
\affiliation{Dept. of Physics, Chung-Ang University, Seoul 06974, Republic of Korea}
\affiliation{Oskar Klein Centre and Dept. of Physics, Stockholm University, SE-10691 Stockholm, Sweden}
\affiliation{Dept. of Physics and Astronomy, Stony Brook University, Stony Brook, NY 11794-3800, USA}
\affiliation{Dept. of Physics, Sungkyunkwan University, Suwon 16419, Republic of Korea}
\affiliation{Institute of Physics, Academia Sinica, Taipei, 11529, Taiwan}
\affiliation{Dept. of Physics and Astronomy, University of Alabama, Tuscaloosa, AL 35487, USA}
\affiliation{Dept. of Astronomy and Astrophysics, Pennsylvania State University, University Park, PA 16802, USA}
\affiliation{Dept. of Physics, Pennsylvania State University, University Park, PA 16802, USA}
\affiliation{Dept. of Physics and Astronomy, Uppsala University, Box 516, SE-75120 Uppsala, Sweden}
\affiliation{Dept. of Physics, University of Wuppertal, D-42119 Wuppertal, Germany}
\affiliation{Deutsches Elektronen-Synchrotron DESY, Platanenallee 6, D-15738 Zeuthen, Germany}

\author[0000-0001-6141-4205]{R. Abbasi}
\affiliation{Department of Physics, Loyola University Chicago, Chicago, IL 60660, USA}

\author[0000-0001-8952-588X]{M. Ackermann}
\affiliation{Deutsches Elektronen-Synchrotron DESY, Platanenallee 6, D-15738 Zeuthen, Germany}

\author{J. Adams}
\affiliation{Dept. of Physics and Astronomy, University of Canterbury, Private Bag 4800, Christchurch, New Zealand}

\author[0000-0003-2252-9514]{J. A. Aguilar}
\affiliation{Universit{\'e} Libre de Bruxelles, Science Faculty CP230, B-1050 Brussels, Belgium}

\author[0000-0003-0709-5631]{M. Ahlers}
\affiliation{Niels Bohr Institute, University of Copenhagen, DK-2100 Copenhagen, Denmark}

\author[0000-0002-9534-9189]{J.M. Alameddine}
\affiliation{Dept. of Physics, TU Dortmund University, D-44221 Dortmund, Germany}

\author[0009-0001-2444-4162]{S. Ali}
\affiliation{Dept. of Physics and Astronomy, University of Kansas, Lawrence, KS 66045, USA}

\author{N. M. Amin}
\affiliation{Bartol Research Institute and Dept. of Physics and Astronomy, University of Delaware, Newark, DE 19716, USA}

\author[0000-0001-9394-0007]{K. Andeen}
\affiliation{Department of Physics, Marquette University, Milwaukee, WI 53201, USA}

\author[0000-0003-4186-4182]{C. Arg{\"u}elles}
\affiliation{Department of Physics and Laboratory for Particle Physics and Cosmology, Harvard University, Cambridge, MA 02138, USA}

% ...

\collaboration{420}{IceCube Collaboration}

\date{\today}

\noaffiliation

\begin{abstract}
    Gravitational-wave events from mergers of compact objects are a predicted source of high-energy neutrinos. Using data from the IceCube Neutrino Observatory, we search for neutrinos coincident with 85 significant and 945 low-significance gravitational-wave candidate events from compact binary coalescences published in real-time by the LIGO-Virgo-KAGRA collaboration during the first part of its fourth observing run (O4a) and its preceding engineering run, within a time window of $\pm500$ seconds from the merger time. We report improvements to the online pipelines, including automatic sending of notices, which improved the median IceCube real-time response compared to the previous observing run by a factor of 2.7 times faster. In addition, we search for long-duration neutrino emission from three candidate events: two neutron star-black hole mergers and one low-significance gravitational-wave event with a possible subthreshold gamma-ray counterpart. We use two methods, both of which have been previously used to search for neutrinos from gravitational-wave transients: an unbinned maximum likelihood analysis on significant alerts and a Bayesian analysis accounting for astrophysical priors on all alerts. We find no statistically significant emission from any individual gravitational-wave event analyzed. In the unbinned maximum likelihood search, the lowest $p$-value event is S231029y, with $p=0.0079$ ($0.752$ post-trials). In the Bayesian search, the lowest $p$-value for significant events is S230904n, with $p=0.0076$ ($0.9996$ post-trials), and for low-significance events is S231025a, with $p=0.0011$ ($0.678$ post-trials), with an overall population $p$-value of $0.55$. We accordingly set upper limits for each event on the time-integrated flux and energy emitted in neutrinos, assuming isotropic emission.
\end{abstract}


\section{Introduction} \label{sec:intro}

Multimessenger observations are a powerful tool to understand fundamental processes occurring in astrophysical sources. Low-latency searches can enable or refine rapid follow-up efforts by multiple telescopes and observatories in response to transient events. The success of multimessenger astronomy is illustrated by the detection of GW170817, a binary neutron star merger, on August 17, 2017 \citep{gw170817_discovery, PhysRevLett.119.161101}. Real-time alerts from observations of the gravitational wave (GW) signal and the subsequent short gamma-ray burst (GRB), GRB 170817A, enabled identification of the merger host galaxy and the detection of an associated kilonova \citep{gw170817_discovery, mma_170817}. 

Following the first detection of astrophysical neutrinos in 2013 \citep{icecube_2013_astro_nu}, the IceCube Neutrino Observatory has identified a flux of astrophysical neutrinos, with evidence deviating from a single power law shape in the diffuse spectrum \citep{icecube-diffuseprl:2025, icecube-diffuseprd:2025}. The 2017 observation of a high-energy neutrino coincident with the blazar TXS 0506+056 initiated a massive follow-up campaign from many telescopes around the globe, which subsequently identified a multi-wavelength flare from that blazar \citep{mm_txs}, illustrating the impact of neutrino astronomy. Other types of AGN have shown evidence of neutrino emission, including the Seyfert galaxy NGC 1068 at the $4.2\sigma$ level \citep{icecube_ngc1068} and excesses from stacking analyses of X-ray bright AGN \citep{IceCube_estes_seyfert, IceCube_xray_seyfert, Abbasi_hard_xray_agn}. Additionally, IceCube has observed $4.5 \sigma$ evidence of neutrino emission from the Galactic plane \citep{icecube_galactic_plane}. Although a few neutrino sources are beginning to emerge, the sources of most of these neutrinos remain unknown.

Compact binary mergers are expected to be a source of high-energy (TeV-PeV) neutrinos \citep{kimura_gwnu, fang_gwnu, Kimura_2017, murase-bartos-review}, and are thus important to follow-up with high-energy neutrino detectors. These events include Binary Neutron Star (BNS), Neutron Star\textendash Black Hole (NSBH), and Binary Black Hole (BBH) mergers\textendash all of which can be detected in GWs by the ground-based GW detector network LIGO \citep{LIGOScientific:2014pky}, Virgo \citep{VIRGO:2014yos}, and KAGRA \citep{KAGRA:2020tym} (LVK) detectors. Relativistic jets formed during BNS or NSBH mergers could accelerate cosmic rays to high energies. The subsequent interactions of these cosmic rays produce pions, which in turn decay to neutrinos \citep{ando_gw_review}. Depending on the viewing angle of the jet, the neutrino expectation can vary \citep{ahlers:2019,biehl:2018}. In the case of BBH events, if those mergers are embedded in sufficiently dense medium, such as an AGN disk, they may also produce neutrinos~\citep{ford2019multimessenger}.

The timescale of neutrino emission in BNS mergers is expected to be similar to that of their short gamma-ray bursts \citep{Baret_2011,Matsui:2023ohr}. Therefore, we select a time window of $\pm 500$ seconds with respect to the merger time for these searches.
Some models also predict longer timescale emission from a millisecond magnetar formed after BNS or NSBH mergers, up to timescales of a few weeks \citep{fang_gwnu}. Therefore, we also search for extended emission with a longer time window of $[-0.1, \, +14]$ days with respect to the merger time.

\begin{figure}
    \centering
    \includegraphics[width=\linewidth]{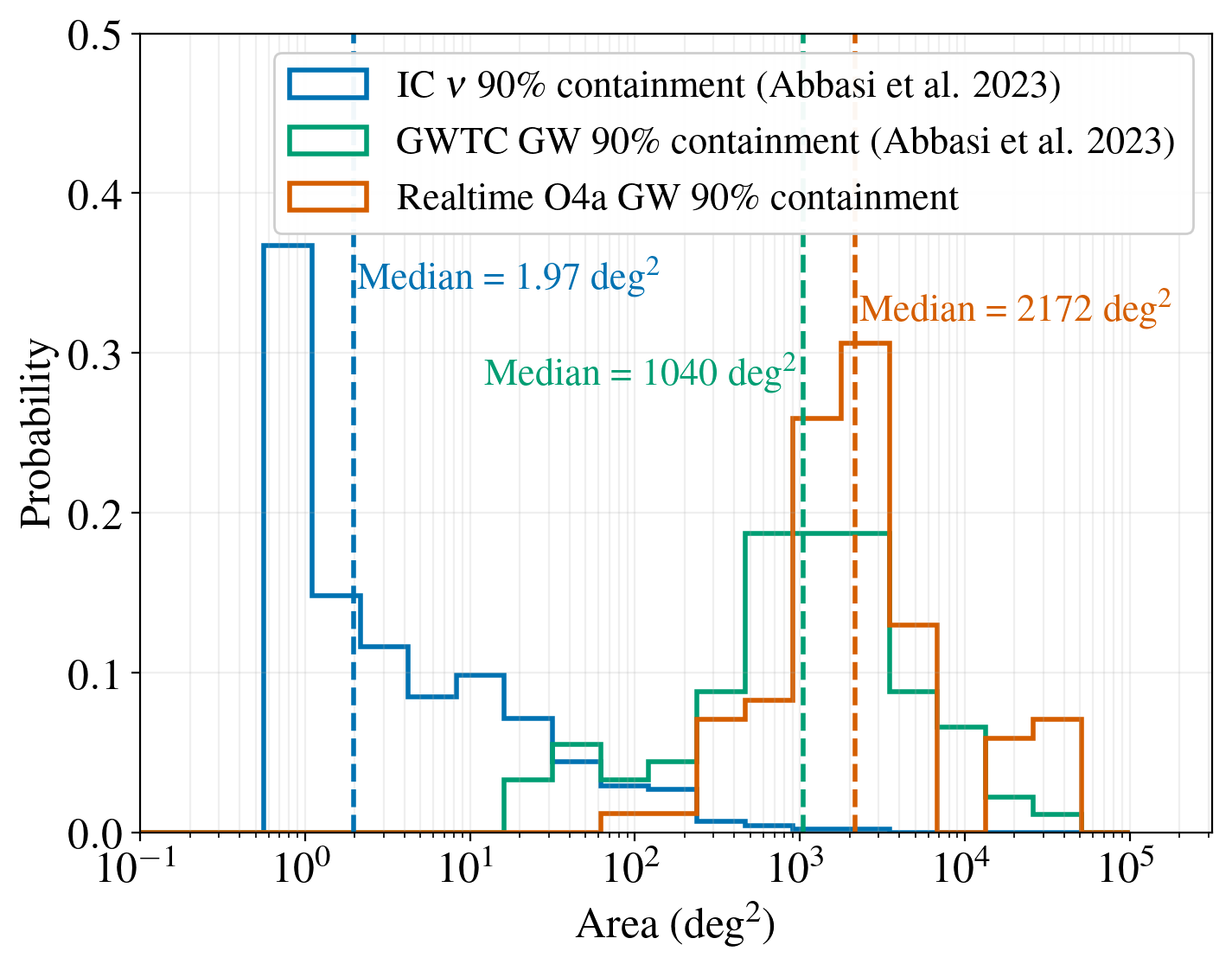}
    \caption{Comparison of 90\% containment areas for neutrino events and GW events. The 90\% area for neutrino events is shown in blue and assumes an $E^{-2}$ source spectrum \citep{icecube_o3}. Previous events reported in the Gravitational Wave Transient Catalog (GWTC, \cite{gwtc1, gwtc2, gwtc2.1, gwtc3}) are shown in green, while candidate events sent publicly during O4a are shown in orange. The median area of the 90\% containment of IceCube events is $10^3$ times smaller than the median 90\% containment area for GW events during O4a.
    }
    \label{fig:area}
\end{figure}

In the past, several searches for neutrino emission from compact binary coalescence events have been performed, although no significant emission has yet been detected \citep{icecube_o3, icecube_o1o2, antares_o2}, including in the case of GW170817 \citep{mma_170817,ic_antares_auger_170817}. Additionally, searches for GeV-scale neutrino emission from interactions of protons with neutrons in the jet \citep{greco_gw} and at lower MeV-scale energies \citep{superk_gw, kamland_gw, borexino_gw} were also carried out. During the previous observing run of the LVK detector network, O3 \citep{PhysRevD.102.062003}, IceCube responded to all GW events sent as public alerts in real time \citep{icecube_o3}. The high uptime of the IceCube detector (>95\%) coupled with its all-sky field of view make IceCube an excellent instrument for follow-up of transients. 
The identification of a neutrino counterpart would help to better understand the physical processes occurring in GW sources, specifically in the relativistic jet \citep{kimura_gwnu}. Additionally, identification of a possible neutrino counterpart in real time can provide a better localization area, often smaller than the GW contour by orders of magnitude, to aid in other searches. Histograms illustrating this comparison of the 90\% containment area for the IceCube dataset used in these searches and for GW events in the published catalog and in the first part of the fourth observing run (O4a) are shown in Fig.~\ref{fig:area}.

In this paper, we report results from low-latency follow-ups of GW events during O4a. The archival analyses on the published catalog are planned for future. We introduce the IceCube Neutrino Observatory and the data sample used in Section~\ref{sec:icecube} and the GW events analyzed in Section~\ref{sec:lvk}. In Section~\ref{sec:methods} we discuss the two methods used for these follow-ups: a generic transient search based on an unbinned maximum likelihood analysis and a Bayesian method employed in the Low-Latency Algorithm for Multimessenger Astrophysics (LLAMA) pipeline. We present updates to IceCube's real-time system for responding to these alerts in Section~\ref{sec:realtime}, and results from the real-time searches in Section~\ref{sec:results}. Finally, we present the conclusions in Section~\ref{sec:conclusion}.

\section{The Neutrino and GW Observatories} 
\subsection{IceCube Detector} \label{sec:icecube}

The IceCube Neutrino Observatory is a cubic kilometer-scale detector instrumented in the ice at the geographic South Pole \citep{icecube_jinst_2017}. It consists of 86 strings, each with 60 digital optical modules (DOMs) instrumented in the ice. DOMs in the main array are spaced 17\,m apart vertically on the strings, with strings spaced 125\,m apart horizontally. This spacing is optimized for the detection of high-energy (TeV\textendash PeV) neutrinos. 

IceCube detects neutrino events via Cherenkov light produced as a result of neutrino interactions with the ice. These events have two main morphologies in the detector, depending on the flavor of neutrino and type of interaction: tracks and cascades. Track events are produced when a muon-flavor neutrino undergoes a charged-current interaction in the ice, while cascade events are produced by electron or tau neutrino charged current interactions or neutral current interactions of all flavors. 
The better angular resolution of track events ($\lesssim 1^\circ$ radius, \cite{icecube_GFU}) compared to cascade events ($\sim10^\circ$ radius, \cite{icecube_galactic_plane}) makes track events the optimal choice for these follow-up searches.

IceCube has established infrastructure to detect and reconstruct neutrino candidate events with low latency to facilitate rapid follow-up of astrophysical transients in real time \citep{realtime:2017, kintscher:2016}. Both of the analyses presented in this paper utilize a dataset of high-energy track-like events available in low-latency from the South Pole, called the ``Gamma-ray Follow-up'' (GFU) dataset \citep{icecube_GFU}. This dataset is optimized for sensitivity to short timescale transients and includes events from the entire sky. The dataset has an all-sky rate between $6-7$~mHz due to seasonal variations of atmospheric backgrounds, shown in \citet[Fig.~1]{IceCube_fra}. This event selection is dominated by atmospheric neutrinos (atmospheric muons) in the Northern (Southern) sky with astrophysical neutrinos making up $\mathcal{O}(0.1\%)$ ($\mathcal{O}(0.01\%)$) of the sample \citep{icecube_GFU, IceCube_fra}. The 90\% central energy range of this dataset ranges from $500$ GeV$-50$ PeV, assuming a source with a spectrum $dN/dE \propto E^{-2}$ and depending on the declination of the source.

\subsection{LVK O4a} \label{sec:lvk}
The first part of the fourth observing run (O4a) of the ground-based GW detectors LIGO, Virgo, and KAGRA (LVK) began on May 24th, 2023, at 15:00 UTC and ran until January 16th, 2024, 16:00 UTC. O4a was preceded by an engineering run (ER15), which started on April 26, 2023, at 15:00 UTC. Information about the observations and detector capabilities in real time are publicly available on the LIGO/Virgo/KAGRA Public Alerts User Guide \citep{lvk-alert-userguide}. There are multiple LVK online search pipelines running in real time to identify GW candidate events during O4a, including 4 pipelines for compact binary coalescence (CBC) events. During O4, public candidate events are separated into two types based on their False Alarm Rate (FAR) by the LVK collaboration\footnote{See \url{https://emfollow.docs.ligo.org/userguide/analysis/index.html\#false-alarm-rate-for-alerts-and-trials-factor} for more details.}. 
Public alerts are sent for events with FAR less than 2 per day, and are designated \textit{significant} by the LVK collaboration if their FAR is less than one per seven months for CBC events or 
one per three years for unmodeled burst searches; otherwise, the event is designated \textit{low-significance}. These FAR thresholds are defined prior to accounting for the number of pipelines online at the time of the alert.
The online automated pipelines detect individual candidate events, and identified candidates associated with a single astrophysical source are aggregated into a single ``superevent'' candidate that is sent publicly~\citep{lvk-alert-userguide}. 
Superevent information and skymap localization for each candidate event is sent over the General Coordinates Network (GCN) \citep{gcn}. For significant events, there is a manual vetting process for each candidate, after which the event is either retracted or an initial notice sent~\citep{lvk-alert-userguide}. 

81 candidate significant GW events were sent in realtime during O4a. All of these are CBC events (no burst events). 
In addition, prior to the start of O4a during the commissioning period for the detectors (ER15), 4 additional candidates were sent publicly~\citep{ER15events}. Although two of these candidate significant events (S230522a and S230522n, both one detector events sent on May 22, 2023) were not manually vetted by LVK, they were not retracted and are thus included in this work. In total, 85 significant events and 945 low-significance CBC candidates are analyzed. Information and skymaps for all events are available from LVK via GraceDB~\citep{gracedb}.

\section{Methods} \label{sec:methods}
\subsection{Unbinned Maximum Likelihood Analysis} \label{sec:uml}
The Unbinned Maximum Likelihood (UML) analysis (also called the \textit{generic transient search} in public GCN Notices) searches for neutrino emission consistent with a point source given a particular GW skymap. It uses an unbinned likelihood method \citep{Braun:2008bg, Braun:2009wp} which has previously been used to search for neutrino point sources, including several analyses in real time \citep{IceCube_fra, IceCube_internal_fra, icecube_o3}. The UML search is used to follow up significant GW alerts from LVK during O4a.

The likelihood is a sum of signal ($\mathcal{S}$) and background ($\mathcal{B}$) PDFs, given by

\begin{align}
    \mathcal{L}(\mathbf{\Omega}, n_s, \gamma)& = \frac{(n_s+n_b)^N e^{-(n_s+n_b)}}{N!} \times \nonumber \\ 
     \prod^N_{i=1} & \left( \frac{n_s}{n_s+n_b} \cdot \mathcal{S}(\mathbf{\Omega}_i, E_i, \sigma_i|\mathbf{\Omega}, \gamma) + \right. \nonumber \\
    &\left. \frac{n_b}{n_s+n_b} \cdot \mathcal{B}(\delta_i, E_i)\right),
\end{align}
where the product iterates over $N$ neutrino events and $n_s,\, n_b$ give the fitted number of signal events and expected number of background events for a location $\mathbf{\Omega}$ in equatorial coordinates. The likelihood takes into account the direction (R.A., decl.) $\mathbf{\Omega}_i=(\alpha_i, \,\delta_i)$, energy $E_i$, and angular resolution $\sigma_i$ of each event $i$. The signal PDF includes a spatial component and an energy component, and assumes a power-law spectrum $F(E)=dN/dE \propto E^{-\gamma}$. The Poisson term in front of the product accounts for the number of neutrino events observed  in the time window (sometimes called an ``extended likelihood'') and is especially important for short timescale analyses due to the low number of events observed. The analysis test statistic (TS) is given by

\begin{equation}
    \rm TS = \max_{\mathbf{\Omega}, n_s, \gamma}\left( 2\ln \left[\frac{\mathcal{L}(\mathbf{\Omega}, n_s, \gamma))}{\mathcal{L}(\mathbf{\Omega}, n_s=0)}\right] +2\ln[w(\mathbf{\Omega)]}\right),
    \label{eq:ts_uml}
\end{equation}
where $w(\mathbf{\Omega})$ is the per-pixel probability of the GW skymap, normalized to the value of the maximum probability pixel of the skymap. We use HEALPix\footnote{https://healpix.sourceforge.io} \citep{healpix_methods} to split the sky into equal-area pixels, and calculate the test statistic in each pixel. We fit a best-fit location $\mathbf{\Omega}$, number of signal events $n_s$ and spectral index $\gamma$ by maximizing the TS.
The $p$-value for each GW event is computed by comparing the unblinded test statistic for a particular map to a background-only test statistic distribution. To make the background-only distribution, we run pseudo-experiments in which the events are scrambled in right ascension and time to preserve the detector acceptance, and calculate background-only test statistics for each pseudo-experiment \citep{IceCube_fra}.

In previous observing runs \citep{icecube_o1o2, icecube_o3}, negative test statistic values were allowed in the analysis. These negative test statistic values occured because the test statistic was only computed for locations within $3\sigma$ of any observed neutrino event. 
For O4, the test statistic is constrained such that $\rm TS \geq 0$, because there is always one location where the penalty term ($2 \ln[w(\Omega)]$ in equation~\ref{eq:ts_uml}) is defined to be 0 (the best-fit location of the GW skymap). This is more consistent with other IceCube real-time follow-ups of astrophysical transients \citep{IceCube_fra, IceCube_internal_fra}.

\subsection{Low-Latency Algorithm for Multimessenger Astrophysics (LLAMA)} \label{sec:llama}
The Low-Latency Algorithm for Multimessenger Astrophysics (LLAMA) pipeline~\citep{countryman2019lowlatencyalgorithmmultimessengerastrophysics, stef_thesis} uses a model-dependent optimal approach~\citep{bartos_2019} to search for high-energy neutrinos coincident with GW events. The model dependence uses our physics and astrophysics knowledge on the generation, propagation, and source properties of GW and high-energy neutrinos via Bayesian priors in the test statistic. The LLAMA search is used to follow up both significant and low-significance CBC GW alerts from LVK.

There are two main reasons for employing model-dependency. First is to use the extra information of luminosity distance provided by the GW observations of CBCs, for which an estimate on the distribution of emitted GW energies or amplitudes is needed. Hence, we use a distribution for the expected GW emission energies, based on past observations. Second, the LLAMA pipeline also considers the fact that not all of the observed GW events are astrophysical. This is even true for confident events. In such a scenario, a uniformly powerful model-independent search for multimessenger events does not exist~\citep{2021ApJ...908..216V}. Every construction of a test statistic implicitly or explicitly favors a model. In LLAMA, we consciously choose which models to favor according to physics and past observations. Coincidences with GW events in O3, including low-significance candidates, were analyzed offline~\citep{o3-subthreshold-llama} with the same method presented here. The results we show in this article include the first real-time high-energy neutrino coincidence analysis of low-significance GW events.

For CBC events, the LLAMA pipeline uses the volume localization of the GW event, sky localization of the high-energy neutrinos, the detection times of the GW event and the neutrinos, the energy proxies of the neutrinos, a proxy for the GW signal to noise ratio, and the $p_{\rm astro}$ value (astrophysical probability) of GW event as inputs ($\mathbf{x}$). The localizations and detection times essentially relate the GW and neutrinos to each other. The remaining inputs are used to quantify the likelihood of GW and neutrinos being astrophysical. We use prior distributions on the location of the astrophysical events (uniform in volume), their reference time (uniform in time), isotropic equivalent emitted GW energy (log uniform distribution in $[10^{-1},10^{1}]~$M$_{\odot}c^2$ considering the released energies in CBCs that have a total mass between 2~M$_\odot$ and 200~M$_\odot$), and isotropic equivalent emitted energy in high-energy neutrinos (log uniform distribution in $[10^{46},10^{51}]$~erg considering the short gamma-ray emission energetics~\citep{2014ARA&A..52...43B} which are thought to be similar to high-energy neutrino emission energetics~\citep{Kimura_2017,biehl:2018,murase-bartos-review}).  We assume the differential number density of the neutrino emission energy spectrum is a power law proportional to $E^{-2}$. The search time window is $\pm500$~s around the GW event, with the test statistic decreasing linearly from the GW detection time to zero at $\pm500$~s, thereby favoring temporally closer coincidences.

To construct our test statistic we define four hypotheses: $H_s$ denotes that the GW and a neutrino are astrophysical and coming from the same source, $H_c^{\rm GW}$ describes that chance coincidence case where the GW is astrophysical but the neutrinos are not, $H_c^{\rm \nu}$ the case where a neutrino is astrophysical but the GW is not, and $H_n$ is the null hypothesis where GW and neutrino triggers are all coming from background. We ignore the extremely rare case of unrelated astrophysical GW and neutrinos. With our priors and detector characteristics, we estimate the prior probabilities of these hypotheses which are proportional to their expected observed rates. Using these prior probabilities we construct our test statistic by weighting the likelihood of each hypothesis as~\citep{bartos_2019}
\begin{multline}
{\rm TS}\left(\mathbf{x}\right)=P\left(\mathbf{x}|H_s\right)P\left(H_s\right)/\left[P\left(\mathbf{x}|H_n\right)P\left(H_n\right)\right.\\\left.+P\left(\mathbf{x}|H_c^{\rm GW}\right)P\left(H_c^{\rm GW}\right)+P\left(\mathbf{x}|H_c^{\nu}\right)P\left(H_c^{\nu}\right)\right].
\label{eq:llamats}
\end{multline}

The background distribution we use for obtaining $p$-values is obtained by creating simulated multimessenger events via injecting fake neutrinos into the GW candidate skymap set. We generate fake neutrino coincidences by scrambling our GFU dataset in time. Analyzing these combinations of GW and neutrinos, we obtain a test statistic distribution coming from coincidences where none of them is a true multimessenger coincidence. We use this one single background distribution for all of the events.

\section{Low-latency Operations} \label{sec:realtime}

Both of these analyses are run in real-time to facilitate low-latency follow-up by other observatories. All results from O4a are sent using GCN as machine-readable Notices over Kafka with the topic \texttt{gcn.notices.icecube.lvk \_nu\_track\_search}\footnote{Documentation hosted at \url{https://gcn.nasa.gov/missions/icecube/} and \url{https://roc.icecube.wisc.edu/public/LvkNuTrackSearch/}}. The schema for this pipeline is hosted on the \texttt{nasa-gcn} GitHub repository\footnote{\url{https://github.com/nasa-gcn/gcn-schema/tree/main/gcn/notices/icecube}}.  

For all notices sent by LVK, IceCube sends a GCN Notice with the results of the search. Additionally, we set operational thresholds for the release of coincident neutrino events, based on the pre-trials analysis $p$-value results from the two pipelines for each event. If either IceCube pipeline identifies a pre-trial analysis $p$-value of $p<0.1$, additional coincident neutrino information is released for individual neutrino events which have a per-event $p$-value less than $0.1$ in the analysis. The coincident neutrino information includes the reconstructed direction and 90\% containment error of the neutrino event, as well as the timing information with reference to the merger time. If either analysis finds a pre-trials analysis $p$-value of $p<0.01$ for a particular GW event, we additionally send the IceCube analysis results as a GCN Circular with the same coincident neutrino information to encourage follow-up by other observatories.
The $p$-values reported in real time do not include trials corrections from multiple hypotheses testing, either from the number of GW events analyzed or the two pipelines run in real time. The false alarm rate of these coincidences can be obtained by multiplying the $p$-values with their corresponding GW trigger rates~\citep{lvk-alert-userguide}.

\begin{figure}
    \centering
    \includegraphics[width=\linewidth]{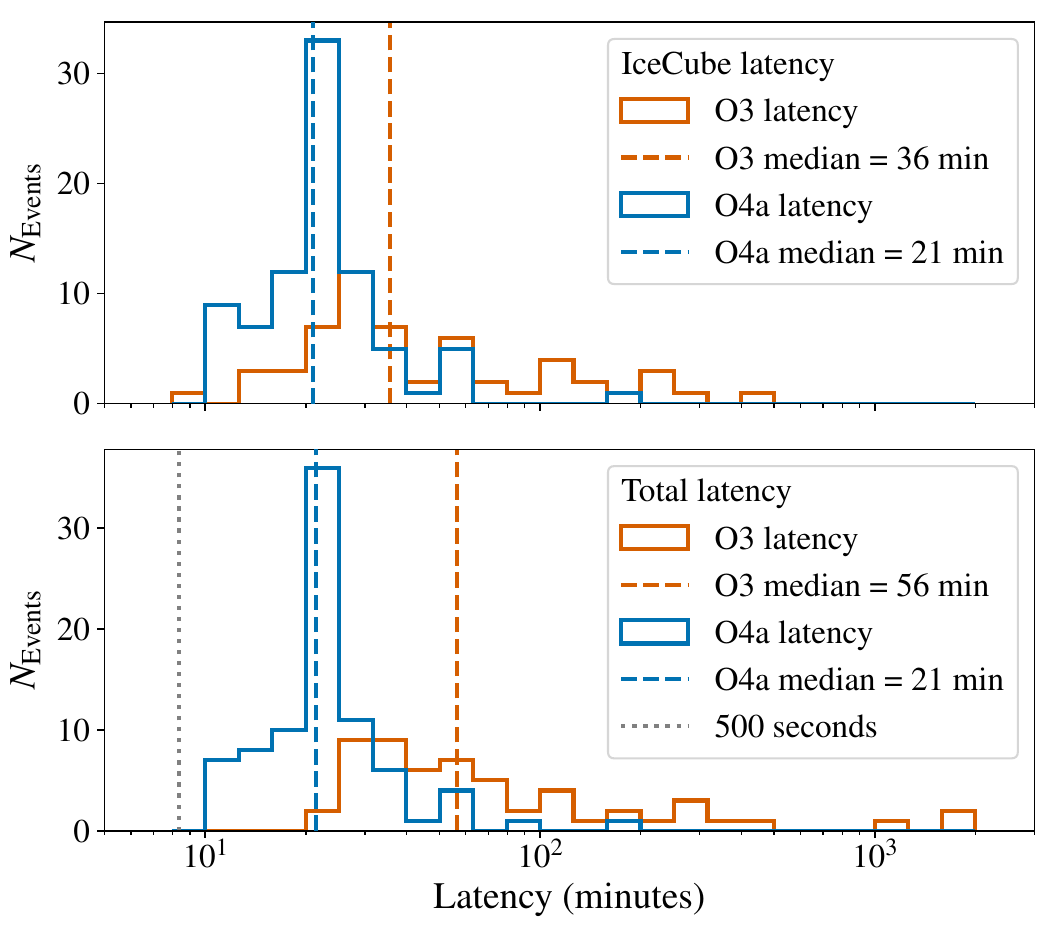}
    \caption{Latency distributions for GW events in O3 and for the first map received for each significant event in O4a. The IceCube latency (top) is the time delay from the receipt of the LVK Notice and publication of the IceCube results GCN, while the total latency (bottom) is defined as the delay between the merger time and publication of the IceCube results. In O4a, most events are sent automatically as GCN Notices, with the exception of events during the engineering run preceding O4a, which were sent as GCN Circulars with human-in-the-loop approval. On the total latency histogram (bottom) 500 seconds after the merger time is plotted (dotted), indicating the end of the analysis time window. Latencies are calculated from publication times of LVK and IceCube public GCN Notices and GCN Circulars \citep{gcn}.}
    \label{fig:latencies}
\end{figure}

A plot of the latency distribution for all significant GW events is shown in Fig.~\ref{fig:latencies}. In the previous observing run, O3, all results sent in realtime were sent via GCN Circulars with human-in-the-loop approval by the IceCube Realtime Oversight Committee (ROC). Because of the increased rate observed for O4, an automated system is used, which ingests results from both pipelines, merges them into a single GCN Notice, and sends it automatically. The improvements in the pipeline, especially the automated sending of results, improved the median latency of the pipeline to be 2.7 times faster with respect to the time of the compact binary coalescence. During ER15 the system was still being set up, so results from the first four events were sent as manual GCN Circulars. Some of these events had higher latencies and thus are outliers in the distribution shown in Fig.~\ref{fig:latencies}. In O4a, the IceCube latency is the dominant contribution to the total latency of the pipeline.

In O4a, there were a total of 13 GCN Circulars (3 significant, 10 low-significance GW events) sent for coincidences with $p<0.01$ in either pipeline identified in real time. One significant event, S230904n \citep{s230904n}, was followed up by other telescopes which identified a potential counterpart in real time. This event was a likely BBH ($p_{\mathrm{BBH}}>99\%$). In real time, the LLAMA pipeline found a $p$-value of 0.004 for the 3-Initial skymap, and coincident event information was circulated in real time \citep{ic230904n}. The direction for the coincident identified event was followed up by the Zwicky Transient Facility, which found a possible counterpart event, AT2023rkw \citep{ztf230904n}. Additional follow-up by the Double Beam Spectrograph on the Palomar telescope identified the event as a Supernova Type 1a, and thus unrelated to the initial GW event \citep{dbsp230904n}. 

\section{Results} \label{sec:results}
Results from the $\pm500$ second analyses with both the LLAMA and UML pipelines, including $p$-values and upper limits on flux and \eiso, are included in Table~\ref{tab:1000s_results} in the Appendix.

\subsection{Results from the UML Search}
First, we present the results from the $\pm500$ second time window searches. For these analyses, all GW events from O1 (September 2015\textendash January 2016), O2 (November 2016\textendash August 2017) \citep{gwtc1}, and O3 (April 2019\textendash March 2020) \citep{gwtc2.1, gwtc3} are included as well as significant events during O4a, for a total of 176 events analyzed. The most significant events for the UML search is the one-detector BBH event S231029y, with a pre-trials $p$-value of $0.0079$. After accounting for 176 candidate events this event does not represent a statistically significant coincidence, with a post-trials $p$-value of $0.752$. We use the \v{S}id\'ak correction method to calculate post-trials $p$-values here and throughout this paper.

\begin{figure}
    \centering
    \includegraphics[width=\linewidth]{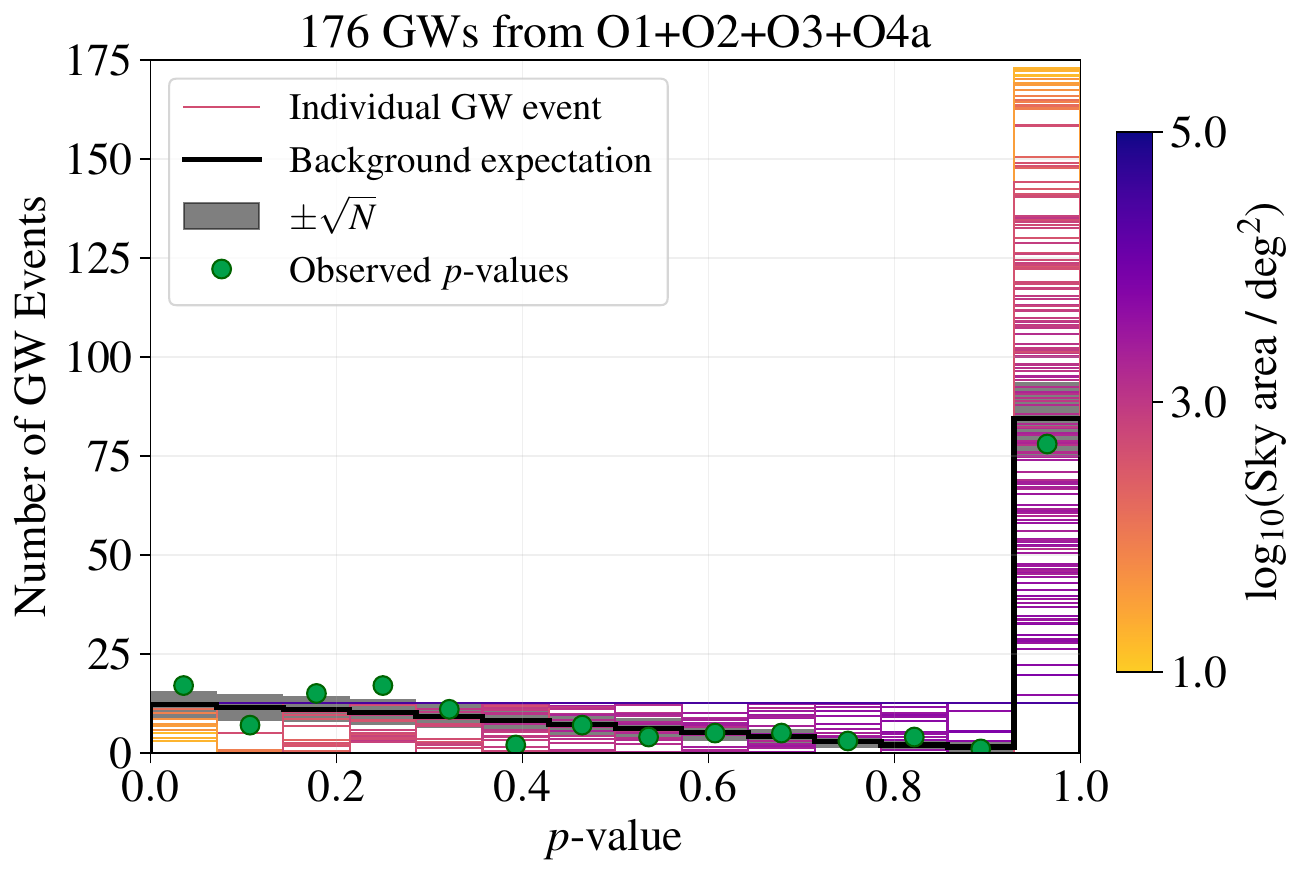}
    \includegraphics[width=\linewidth]{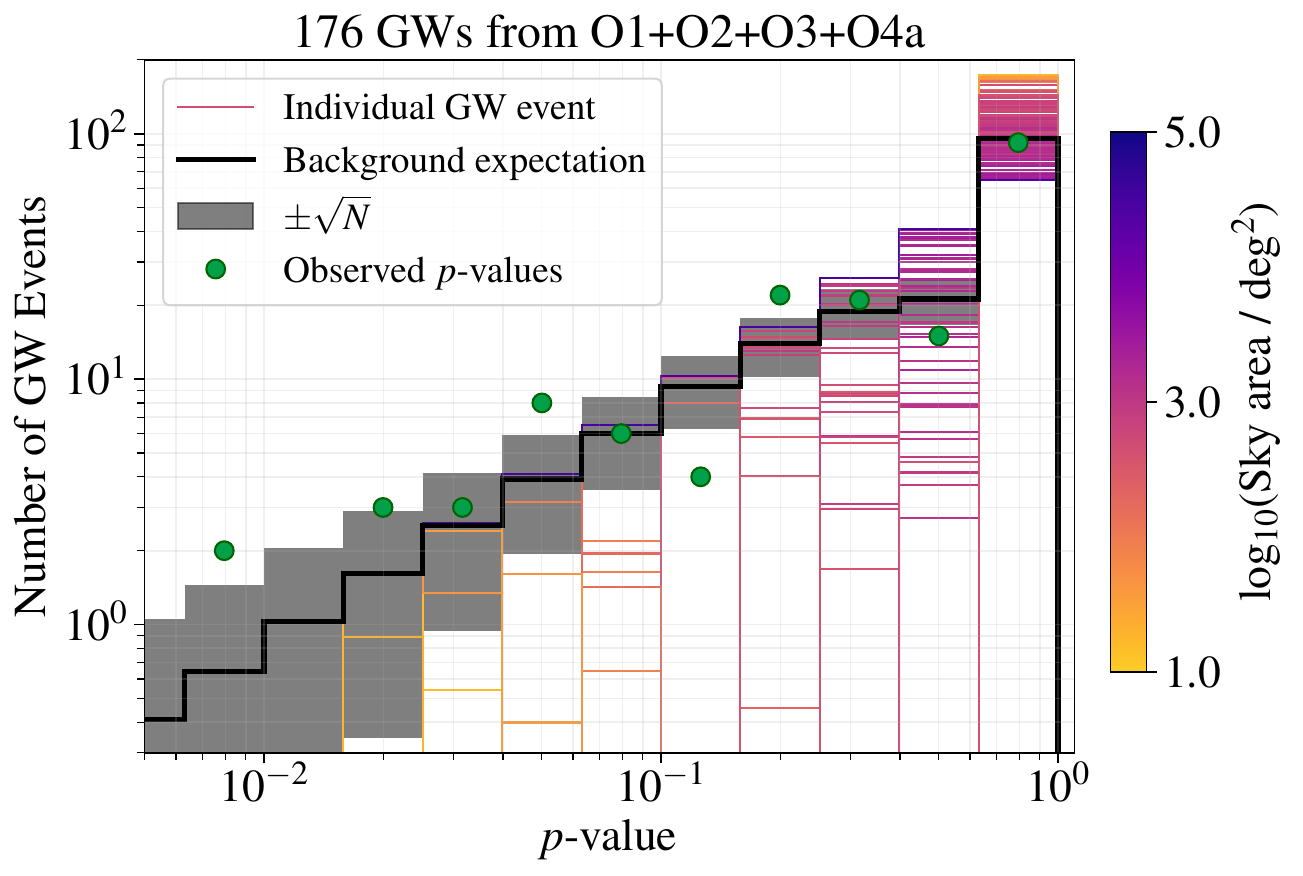}
    \caption{Histogram of $p$-values from the UML search for GW events during O1-O4a. Observed $p$-values in each bin are shown as green points. The background expectation for the ensemble of GW events is shown in the black line, with the Poisson uncertainty ($\sqrt{N}$) indicated by the gray band. The background-only expectation for each GW event is shown in the colored histograms, with the color representing the 90\% containment area of that event’s skymap. There are 30,000 background-only pseudo-experiments for each GW event, as shown in Fig.~\ref{fig:uml_bkg_ts}. The distribution is shown with linear bins (top) and log-scaled bins (bottom). 
    }
    \label{fig:uml_pval}
\end{figure}

The $p$-value distributions for the UML search are shown in Fig.~\ref{fig:uml_pval}. The observed $p$-value distribution is consistent with the background-only distribution within errors. For individual GW skymaps, the background-only $p$-value distribution changes due to the number of expected background coincidences which grows with skymap area.

\begin{figure}
    \centering
    \includegraphics[width=\linewidth]{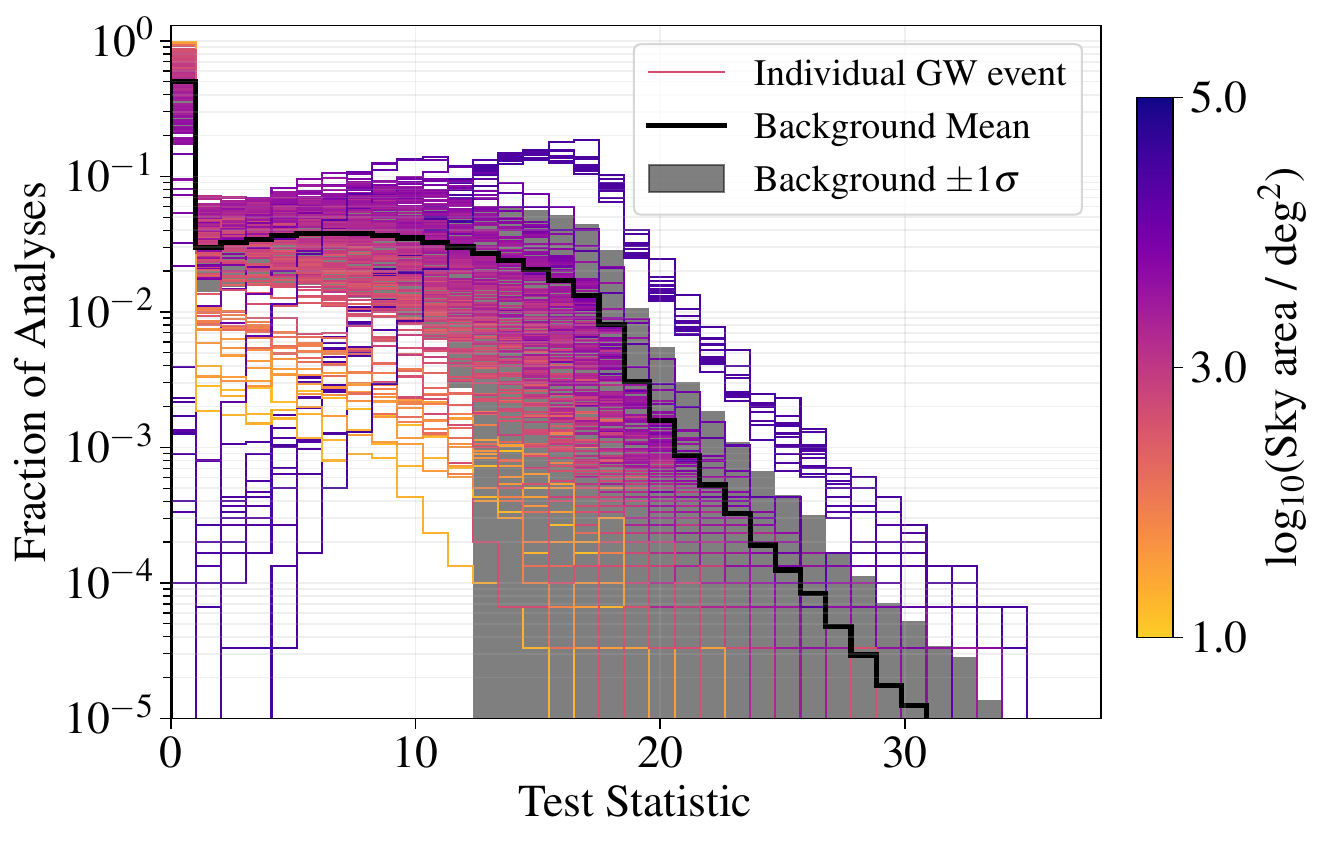}
    \caption{Background-only test statistic distribution for 176 GW events for O1-O4a in the UML search. Each individual GW event is shown in the thin colored lines, corresponding to the map 90\% containment area. Events with smaller skymap areas have lower probability of background coincidences and thus have a higher number of trials with $\rm TS=0$, while events detected in only one GW detector have a very large skymap which causes the peak of the distribution to shift to higher test statistic values. The mean (standard deviation) over all distributions is shown in the black line (gray band).}
    \label{fig:uml_bkg_ts}
\end{figure}

The shape of the background-only $p$-value distribution can be understood by studying the variation in the expected background test statistic distribution for individual GW events, shown in Fig.~\ref{fig:uml_bkg_ts}. The size of the GW skymap for each event changes the shape of the background distribution, as a larger skymap has more expected coincidences from background. One detector events, which have very large skymaps covering roughly half of the sky at 90\% containment, have the highest expected coincident background rate; thus, the background-only distributions shift to higher values of the test statistic. Smaller skymaps with area $\sim10-100$ deg$^2$ have more background-only pseudo-experiments with $\rm TS=0$ corresponding to no events spatially coincident with the skymap. This variation impacts the mean and standard deviation of the background (gray band in Fig.~\ref{fig:uml_bkg_ts}), as there is larger variation in the background distribution for the individual skymaps in these bins. 

The effect of these differences on the background-only $p$-value distribution (Fig.~\ref{fig:uml_pval}) can be seen in the size of the $p=1.0$ bin, as well as the shape of the distribution as a whole. Background pseudo-experiments with $\rm TS=0$ correspond to $p=1.0$, thus for events with small skymaps the background-only $p$-value distribution is more bimodal, with a larger $p=1.0$ peak. For those with large skymaps, the distribution is more uniform. This variation in shape causes the standard deviation in each bin to increase as $p$-value increases.

\begin{figure}
    \centering
    \includegraphics[width=\linewidth]{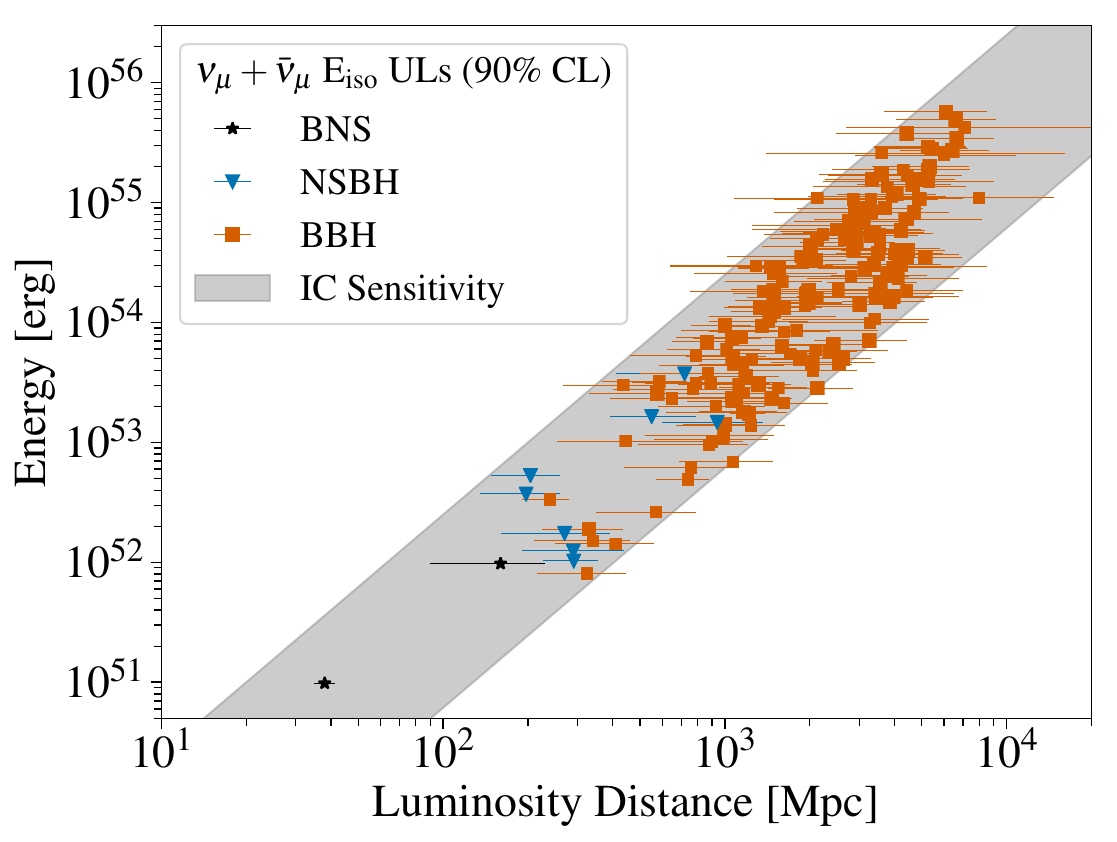}
    \caption{\eiso~90\% CL $\nu_\mu+\bar{\nu}_\mu$ upper limits for each significant GW alert in runs O1-O4a. The gray band shows the IceCube point source sensitivity over all declinations for a $\pm500$ second time window. Distances and distance uncertainties for each event from \cite{gwtc1, gwtc2, gwtc2.1, gwtc3} and public FITS headers for O4a are shown. For GW170817, the distance uncertainty of the host galaxy is used \citep{gw170817_distance}. $\mathrm{E_{iso}}$ values are computed over the 90\% central energy range (500 GeV - 50 PeV), assuming a spectrum $F=dN/dE \propto E^{-2}$. }
    \label{fig:eiso}
\end{figure}

The observed test statistic for each GW event is also used to set limits on the energy emitted in high energy neutrinos, assuming isotropic emission, (\eiso), shown in Fig.~\ref{fig:eiso}. The \eiso~is related to the neutrino flux according to

\begin{equation}
    E_\mathrm{iso} = 4\pi r^2 \Delta T\int_{E_1}^{E_2} E\cdot F(E) \,dE,
\end{equation}
where the neutrino flux $F(E)=dN/dE \propto E^{-2}$, $\Delta T$ is the time window ($\pm 500$ seconds), and $r$ is the source distance. To calculate the \eiso~upper limit, we run pseudo-experiments which sample pixels from the skymap according to their given probability, and sample from that pixel's distance posterior following the procedure in \cite{Singer_distance_supplement}. Using these sampled values, we inject a flux $F(E)$ at the sampled declination for the particular \eiso~value calculated from the skymap. We then compute the upper limit on \eiso, which is marginalized over the skymap in both distance and location for each GW event.
These \eiso~upper limits are only presented for the UML search and not the LLAMA search because the limits would be very similar between the two analyses (see Table~\ref{tab:1000s_results} for all upper limits with both analyses). The bounds on the integral are the 90\% central energy range of the GFU dataset for this spectrum, taken to be $E_1=500$ GeV and $E_2=50$ PeV, to focus on the energy range where IceCube has the best sensitivity. In previous analyses \citep{icecube_o1o2,icecube_o3} the full MC energy range was used for the bounds of the integral ($E_1=10$ GeV, $E_2=10^{9.5}$ GeV), resulting in a difference of a factor of 1.7 compared to previous limits. We rescale the limits from O1, O2, and O3 by this factor to be comparable with limits calculated during O4a. 

In addition to the $\pm500$ second analyses, events which have high probability of at least one neutron star progenitor are analyzed with the extended time window of $[-0.1,\, +14]$ days. An event must pass at least one of three criteria to be analyzed in this search window: 
\begin{enumerate}
    \item $p_{\mathrm{BNS}}+p_{\mathrm{NSBH}} >0.5$;
    \item $\texttt{HasNS} > 0.5$; ($\texttt{HasNS}$ is defined as the probability that one of the objects has a mass consistent with a neutron star \citep{lvk-alert-userguide})
    \item $\texttt{HasRemnant}>0.5$ ($\texttt{HasRemnant}$ is defined as the probability that a nonzero mass was ejected during the merger of the two objects \citep{lvk-alert-userguide}).
\end{enumerate}

Two events, S230518h \citep{s230518h}
and S230529ay \citep{s230529ay}, during O4a passed these criteria. Although another significant event, S230627c \citep{s230627c}, had a $p_{\mathrm{NSBH}}=0.49$, it did not pass any of the three \textit{a priori} defined criteria to be analyzed with this time window, thus it is not included in the two-week analysis.

In the extended time window analyses the expected number of background events is higher than for the shorter time window analyses. The expected number of background events is declination-dependent, and the declination of the highest probability regions of the skymap also impacts the expected background, which limits our sensitivity. The IceCube sensitivity using the skymap for S230518h (S230529ay) in the $\pm500$ second window is $0.49~\mathrm{GeV~cm}^{-2}$  ($0.11~\mathrm{GeV~cm}^{-2}$), while in the 2 week time window, it is $1.04~\mathrm{GeV~cm}^{-2}$ ($0.82~\mathrm{GeV~cm}^{-2}$). This sensitivity is $2.1$ times higher for S230518h and $7.5$ times higher for S230529ay. S230529ay is a single-detector event and thus has a very large skymap (the 90\% containment area for this skymap is more than half the sky), meaning that the expected background is much larger for the longer time window for this map, as indicated by the loss in sensitivity.

There was also a low-significance GW event, S230619bd, which had a subthreshold gamma-ray burst counterpart identified by the Fermi-GBM Targeted Search Pipeline \citep{gbm230619}. Due to this potential coincidence, we analyzed this event with the extended time window in real time using the joint localization map provided by the GBM team \citep{gbm_2023_joint}. 

Results from all three follow-ups are shown in Table~\ref{tab:2week_results}. No significant emission is observed in any of the follow-ups with the extended time window, so we set upper limits on neutrino flux emitted in each event for this time window. Including these three events, there have been 11 events from O2 through O4a analyzed with this window to date (see \cite{icecube_o3, ic_antares_auger_170817} for events analyzed in previous observing runs).

\begin{deluxetable}{cccc}[t!]
\tablecaption{Table of results for the unbinned maximum likelihood analysis with an analysis time window of $[-0.1,\, +14]$ days. All reported $p$-values are pre-trials, and do not account for the number of GW events analyzed. $E^2F$ upper limits are given at the 90\% confidence level (CL). \label{tab:2week_results}}
\tablehead{
\colhead{Event} & 
\colhead{Pre-trial} & \colhead{$E^2F$ 90\% UL} & \colhead{IceCube GCN}\\[-0.2cm]
& \colhead{$p$-value} & \colhead{(GeV cm$^{-2}$)} & \colhead{Circular}\\[-0.4cm]
}
\startdata
S230518h	& 0.754 	&1.04 	&\gcnlink{33907} \\ 
S230529ay 	& 0.394 	&0.94 	&\gcnlink{33980} \\ 
S230619bd/GBM-230619	& 0.343 	&0.10 	&\gcnlink{34142} \\ 
\enddata
\end{deluxetable}

\subsection{Results from the LLAMA search}
\begin{figure}
    \centering
    \includegraphics[width=\columnwidth]{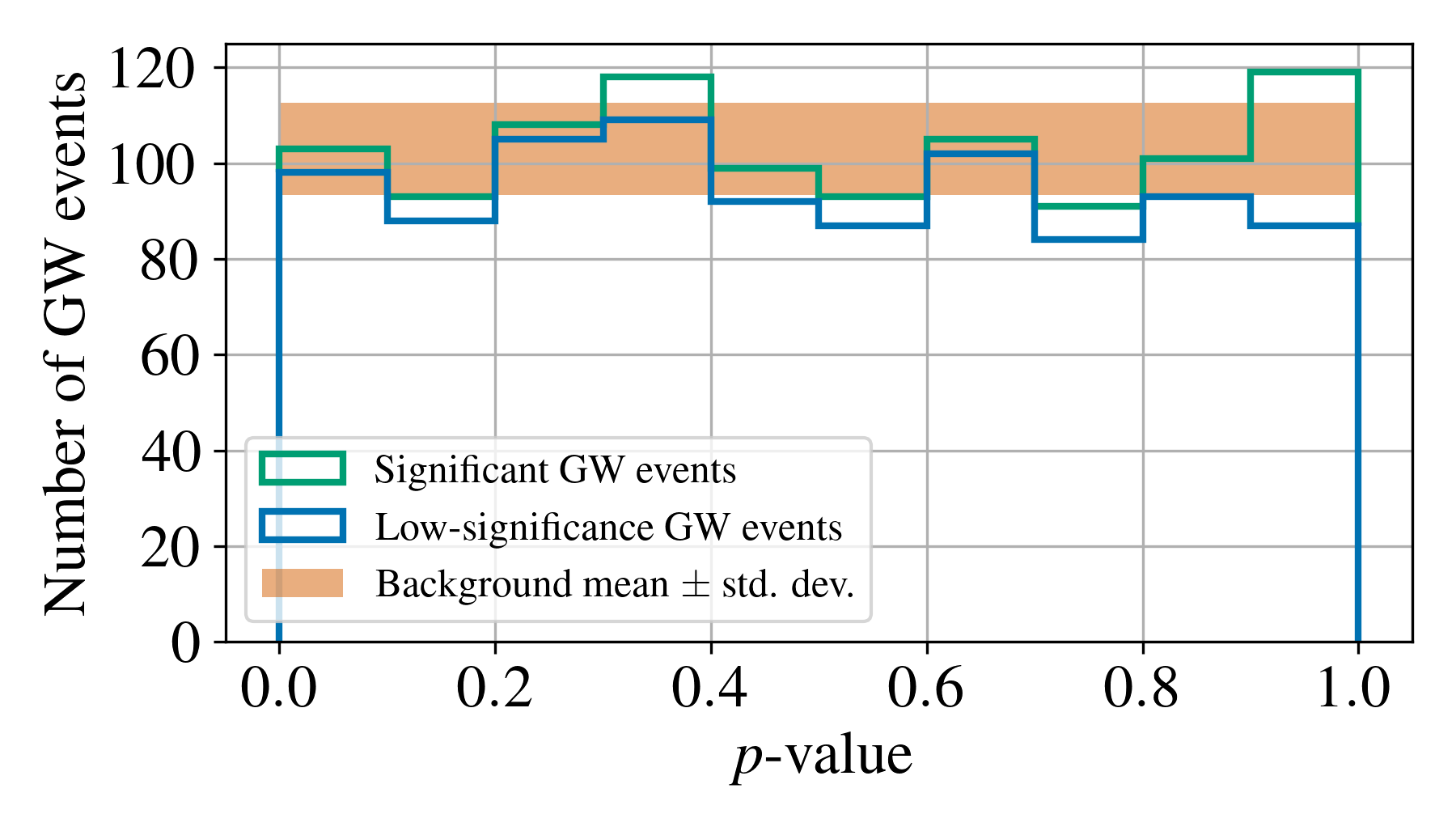}
    \includegraphics[width=\columnwidth]{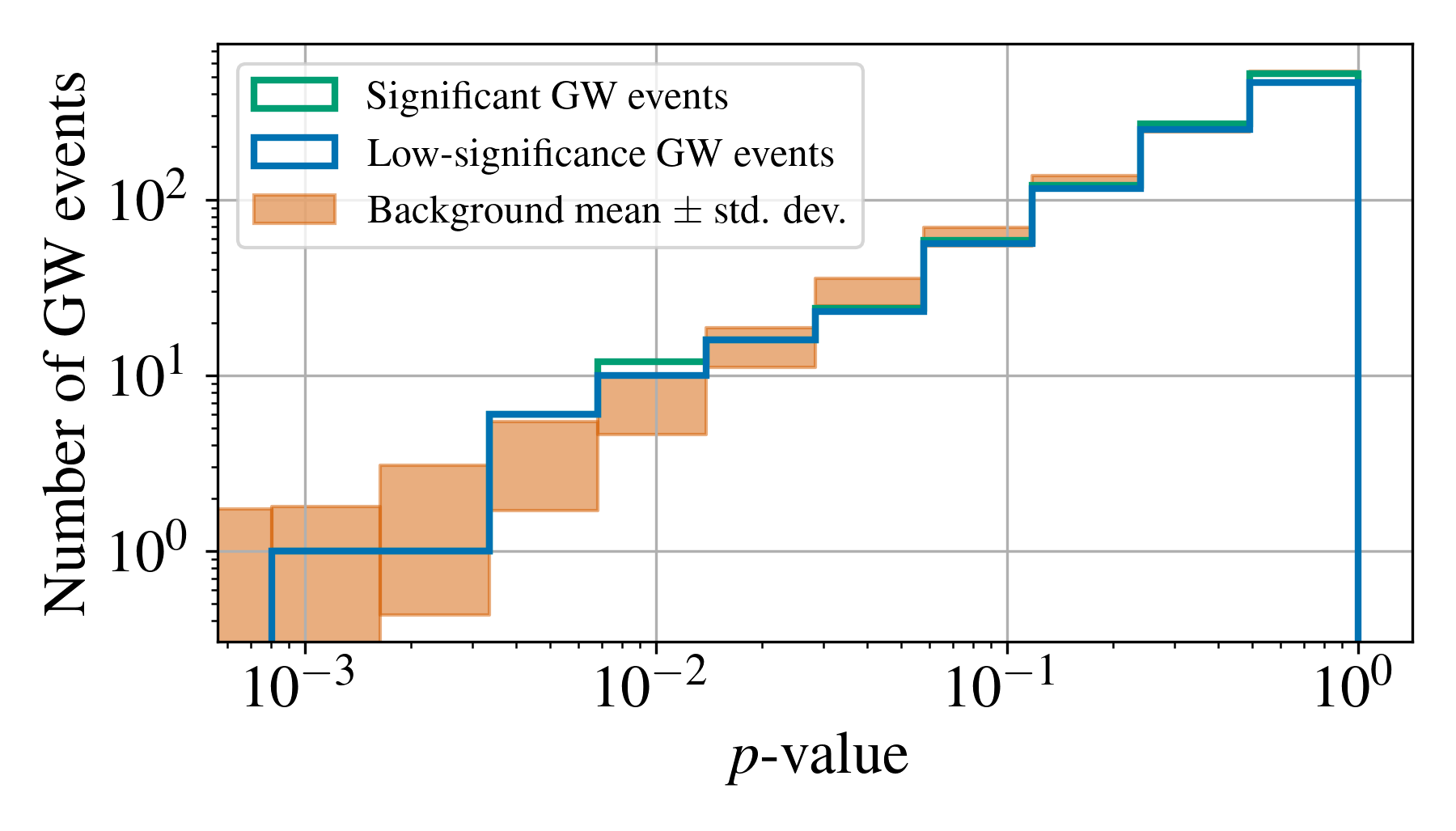}
    \caption{The $p$-value distribution of the results of the LLAMA pipeline, with linear (top) and logarithmic (bottom) bins. The $p$-value distributions for the coincidences of significant and low-significance GW events are distinguished and stacked on top each other. The expected distribution of the background $p$-values is a uniform distribution. The error bands represent binomial uncertainties. The leftmost error band on the bottom figure is for the bin starting from 0.}
    \label{fig:llama_pval}
\end{figure}

The results of the LLAMA search are collectively summarized with the $p$-value distribution shown in Fig.~\ref{fig:llama_pval}. The observed distribution is consistent with a uniform distribution according to background expectation, showing no significant deviation. To quantify this, we construct a new test statistic which is the sum of individual test statistics (in Eq. \eqref{eq:llamats}) ${\rm TS^{pop}}=\sum_i {\rm TS}_i$. After constructing a corresponding background distribution for it, the population $p$-value obtained is 0.55, in agreement with Fig.~\ref{fig:llama_pval}. 

\begin{figure}
    \centering
    \includegraphics[width=\columnwidth]{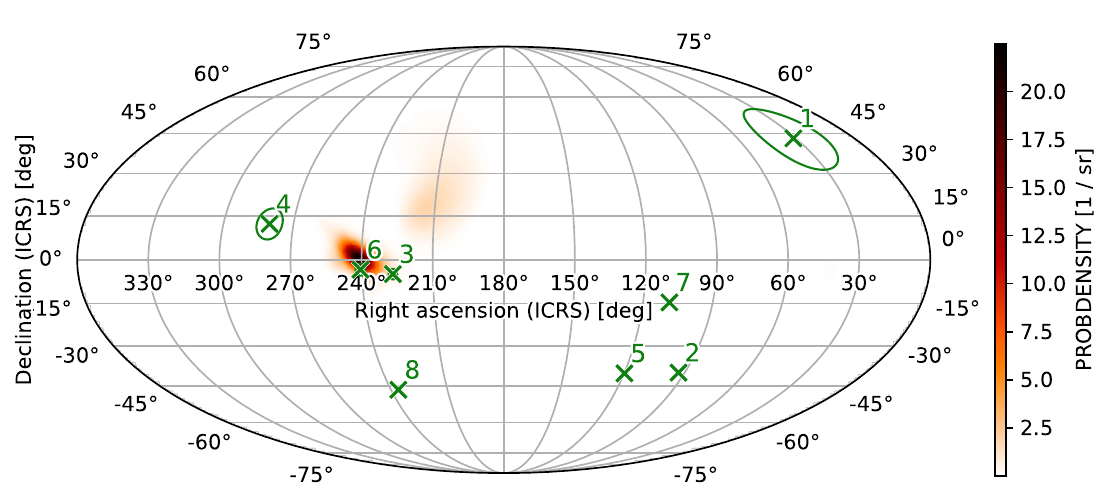}
    \caption{The neutrino coincidences within the $\pm500$~s of the plausible BNS merger trigger S231025a, producing the most significant coincidence found during O4a.}
    \label{fig:lowestpval-llama}
\end{figure}

The lowest $p$-value among the individual events belongs to the low-significance GW candidate event S231025a with a pre-trials $p$-value of $0.0011$. We show the localizations of the candidate GW event and the neutrinos within the $\pm500$~s time window in Fig.~\ref{fig:lowestpval-llama}. Most of the significance comes from the neutrino \#6 which has an energy of $\sim1$~TeV. The coincidence with neutrino \#3 also has a mildly low $p$-value of 0.025. The GW trigger has a false alarm rate of about 30 per year. The source is estimated to be a BNS merger with 59\% probability, and noise with 41\% probability. Assuming it is astrophysical, the BNS merger distance is estimated as $233\pm135$~Mpc, which is on the upper end of LIGO detectors' O4a range sensitivity. This explains the low signal-to-noise ratio of the event. After correcting for trials from a total of 1030 analyzed events, there is no evidence for a neutrino counterpart to the S231025a GW candidate event, with a post-trials $p$-value of $0.678$. Among the confirmed significant LVK alerts, the coincidence with the lowest $p$-value belongs to S230904n with a pre-trial $p$-value of $0.0076$ (post-trials $p$-value of $0.9996$ after accounting for 1030 events). We conclude there is no evidence for inferring a true coincidence after accounting for trials. We find 90\% neutrino fluence upper limits on significant GW candidate events assuming an emission spectrum of a power law with index $-2$. The results for significant GW candidate events are listed in Table~\ref{tab:1000s_results}. The LLAMA results for low-significance candidate events are available on Dataverse\footnote{Data release available on Dataverse: \url{https://doi.org/10.7910/DVN/E0B2OW}} as an accompanying data release.

\subsection{Comparison of results in both pipelines}
\begin{figure}
    \centering
    \includegraphics[width=\linewidth]{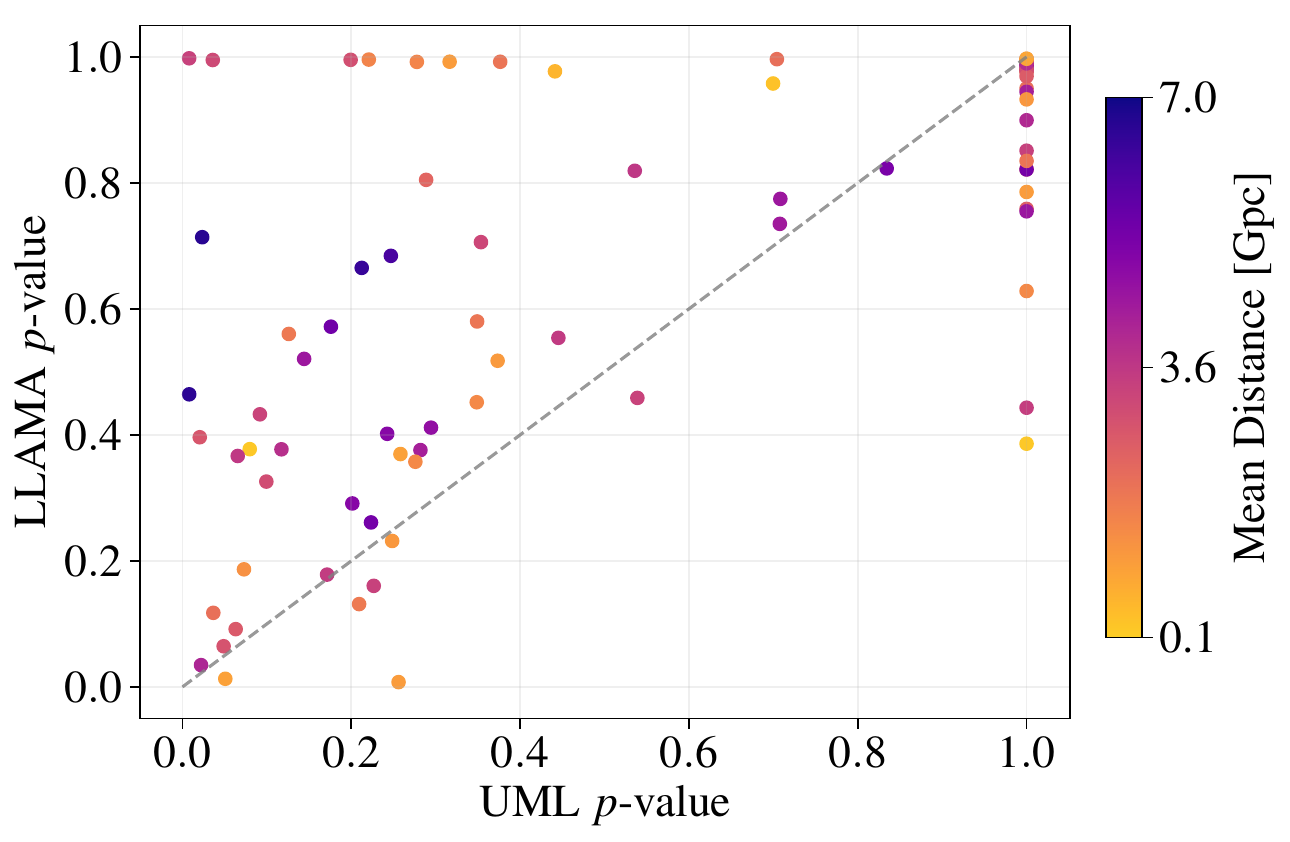}
    \caption{Comparison of pre-trials $p$-values of the significant alerts from the two analysis pipelines for events during O4a. The color scale indicates the mean of the distance posterior to a given GW event for the most recent skymap. 
    }
    \label{fig:pval_comparison}
\end{figure}

A comparison of $p$-values for each individual analysis in the two pipelines is shown in Fig.~\ref{fig:pval_comparison}. Differences in the analysis results are expected due to differences in the methods and priors considered. In particular, the prior on source distance is included for the LLAMA search, indicated in the color bar in Fig.~\ref{fig:pval_comparison}, but is not included for the UML search. The line-like feature where $p_{\mathrm{UML}}=1.0$ on the right side of this figure are those which find $\rm TS=0$ in the UML analysis. The two events in the upper left corner of the plot are S231029y and S230522a, both BBH events detected by a single GW detector. S231029y had a distance of $3.3 \pm 1.3$~Gpc, which had a high energy neutrino in the Southern sky near the 90\% contour of the GW event. The high energy of the event contributes to the low UML $p$-value, while the far distance of the event increases the LLAMA $p$-value.

\section{Conclusion and Outlook} \label{sec:conclusion}
We report no significant neutrino emission from compact object mergers observed during the O4a observation period of LVK. We set upper limits on the flux of neutrinos emitted in a $\pm 500$ second window surrounding the merger time for each individual significant event for both the LLAMA and UML analysis methods. In addition, we report no significant neutrino emission in the extended $[-0.1,\, +14]$ day window from two NSBH mergers during O4a and one reported possible subthreshold GRB counterpart to a low-significance GW event using the unbinned maximum likelihood search. We also set limits on the energy emitted in high energy neutrinos assuming isotropic emission for each significant event. The addition of these 85 events detected during O4a have nearly doubled the number of candidate GW events analyzed with these searches compared to the 91 GW events previously analyzed \citep{icecube_o1o2, icecube_o3}. Importantly, as the LVK collaboration is now releasing public low-significance candidates in real time during O4, the analysis of these events can be done in real time rather than in archival searches. Rapid follow-up of these events in real-time is extremely valuable, as identification of a counterpart can increase the significance of a candidate and inform additional follow-up \citep{o3-subthreshold-llama}.

These analyses were performed in real time on all events during the O4 run. During O4a, the Virgo detector was absent, but Virgo has rejoined the GW detector network for O4b and O4c, leading to several well-localized three-detector events. Thus far, there have been no detected significant BNS merger events during O4, but the planned intermediate run, IR1, is expected to provide an additional 6 months of observing time\footnote{See~\cite{observing-plan} for the most up-to-date information about LVK observing run plans.}. IceCube analysis of events in the GWTC-4.0 \citep{gwtc4, LvkGW230529} that are not included here as well as events from O4b and O4c that are added to the next GWTC releases are also planned for a future publication.

Improvements to the automatic selection of events by LVK has allowed most candidate events to be published to the community via GCN roughly 30 seconds after the time the compact objects merge. The median latency for receiving GFU events from the IceCube detector is $\sim 30$ seconds \citep{realtime:2017, raamis_thesis}, providing a lower bound on the possible latency with this IceCube high energy tracks dataset. The selected time window for these searches ends $+500$ seconds after the merger time, meaning a significant portion of the total latency for the IceCube response is due to waiting for the end of this time window. Modeling of neutrino production in mergers indicates that the neutrino emission is expected to occur on shorter timescales than the current window \citep{Kimura_2017, biehl:2018}. Although reducing this time window would not significantly impact the sensitivity of these searches for well-localized events, using a shorter time window would reduce the overall latency in sending the IceCube results, thus improving prospects for identifying a possible joint source in the case of a significant coincidence.

In the future, improvements planned for the GW detectors will improve the sensitivity of the GW detectors to merger events, increasing the rate of detected events and the expected distance horizon for all GW events. The planned expansion of IceCube-Gen2 to eight times IceCube's current instrumented volume will improve the sensitivity to these events by over an order of magnitude \citep{Gen2TDR}. With the improved effective area of IceCube-Gen2, neutrino-GW coincidences from merger events are more likely to be detected \citep{Mukhopadhyay:2023niv, Matsui:2023ohr}.

\section*{Acknowledgments}
The IceCube Collaboration acknowledges the significant contributions to this manuscript from Jessie Thwaites and Do\u{g}a Veske. This material is based upon work supported by NSF's LIGO Laboratory which is a major facility fully funded by the National Science Foundation.
The authors gratefully acknowledge the support from the following agencies and institutions:
USA {\textendash} U.S. National Science Foundation-Office of Polar Programs,
U.S. National Science Foundation-Physics Division,
U.S. National Science Foundation-EPSCoR,
U.S. National Science Foundation-Office of Advanced Cyberinfrastructure,
Wisconsin Alumni Research Foundation,
Center for High Throughput Computing (CHTC) at the University of Wisconsin{\textendash}Madison,
Open Science Grid (OSG),
Partnership to Advance Throughput Computing (PATh),
Advanced Cyberinfrastructure Coordination Ecosystem: Services {\&} Support (ACCESS),
Frontera and Ranch computing project at the Texas Advanced Computing Center,
U.S. Department of Energy-National Energy Research Scientific Computing Center,
Particle astrophysics research computing center at the University of Maryland,
Michigan State University,
Astroparticle physics computational facility at Marquette University,
NVIDIA Corporation,
and Google Cloud Platform;
Belgium {\textendash} Funds for Scientific Research (FRS-FNRS and FWO),
FWO Odysseus and Big Science programmes,
and Belgian Federal Science Policy Office (Belspo);
Germany {\textendash} Bundesministerium f{\"u}r Forschung, Technologie und Raumfahrt (BMFTR),
Deutsche Forschungsgemeinschaft (DFG),
Helmholtz Alliance for Astroparticle Physics (HAP),
Initiative and Networking Fund of the Helmholtz Association,
Deutsches Elektronen Synchrotron (DESY),
and High Performance Computing cluster of the RWTH Aachen;
Sweden {\textendash} Swedish Research Council,
Swedish Polar Research Secretariat,
Swedish National Infrastructure for Computing (SNIC),
and Knut and Alice Wallenberg Foundation;
European Union {\textendash} EGI Advanced Computing for research;
Australia {\textendash} Australian Research Council;
Canada {\textendash} Natural Sciences and Engineering Research Council of Canada,
Calcul Qu{\'e}bec, Compute Ontario, Canada Foundation for Innovation, WestGrid, and Digital Research Alliance of Canada;
Denmark {\textendash} Villum Fonden, Carlsberg Foundation, and European Commission;
New Zealand {\textendash} Marsden Fund;
Japan {\textendash} Japan Society for Promotion of Science (JSPS), Ministry of Education, Culture, Sports, Science and Technology (MEXT), and Institute for Global Prominent Research (IGPR) of Chiba University;
Korea {\textendash} National Research Foundation of Korea (NRF);
Switzerland {\textendash} Swiss National Science Foundation (SNSF).
\software{ 
\texttt{Astropy}~\citep{astropy:2013, astropy:2018, astropy:2022},
\texttt{gcn-kafka}~\citep{gcn-kafka},
\texttt{icecube/FastResponseAnalysis}~\citep{IceCube_fra},
\texttt{HEALPix}~\citep{healpix_methods},
\texttt{healpy}~\citep{Zonca2019},
\texttt{ipython}~\citep{perez2007ipython},
\texttt{LLAMA}~\citep{countryman2019lowlatencyalgorithmmultimessengerastrophysics},
\texttt{matplotlib}~\citep{hunter2007matplotlib}, 
\texttt{mhealpy}~\citep{2022AJ....163..259M},
\texttt{NumPy}~\citep{harris2020array}
\texttt{pandas}~\citep{pandas, mckinney-proc-scipy-2010},
\texttt{pygcn}~\citep{pygcn},
\texttt{SciPy}~\citep{2020SciPy-NMeth}.
}

\bibliography{bib}

\appendix
\restartappendixnumbering
\section{Table of Results} \label{app:results}
All results for both pipelines for all significant GW events in the $\pm500$ second time window in O4a are tabulated in Table~\ref{tab:1000s_results}. Parameters for each GW event are given, including most likely event type, 90\% containment area of the GW skymap, and distance to the event. Analysis $p$-values included in the table are not trials corrected for the number of GW sources analyzed. 90\% CL upper limits on the flux for each pipeline are included, as well as 90\% CL upper limits on the energy emitted in high energy neutrinos, assuming isotropic emission (\eiso).

\startlongtable
\begin{deluxetable*}{ccccccccccc}
\tablecaption{Table of results for individual significant GW events in each analysis, using a time window of $\pm 500$ seconds. Upper limits on both flux and \eiso~are calculated assuming a source spectrum $F=dN/dE \propto E^{-2}$. \eiso~upper limits are calculated using the 90\% central energy range of the dataset ($500$~GeV$ - 50$~PeV) as the limits of the integration.}

\tablehead{
\multicolumn{5}{c}{GW Parameters} & \multicolumn{3}{c}{LLAMA} & \multicolumn{3}{c}{UML} \\
\cline{1-4}
\cline{6-7}
\cline{9-11}
\colhead{Event} & \colhead{Type} & \colhead{90\% Area} & \colhead{Distance} & & \colhead{Pre-trial} & \colhead{$E^2F$ 90\% UL} && \colhead{Pre-trial} & \colhead{$E^2F$ 90\% UL}  & \colhead{$E_{\rm iso}$ 90\% UL} \\ [-0.2cm]
& & \colhead{(deg$^2$)} & \colhead{(Mpc)} & & \colhead{$p$-value} & \colhead{(GeV cm$^{-2}$)} &  & \colhead{$p$-value} & \colhead{(GeV cm$^{-2}$)} & \colhead{(erg)}
}

\startdata
S230518h	& NSBH	& 460	& 204 $\pm$ 57 &  & 0.386 & 0.437 && 1.000 & 0.488 &  5.31$\times 10^{52}$\\ 
S230520ae	& BBH	& 1702	& 2014 $\pm$ 663 &  & 0.822 & 0.479 && 1.000 & 0.458 &  4.33$\times 10^{54}$\\ 
S230522a${}^\star$	& BBH	& 24218	& 3102 $\pm$ 1032 &  & 0.995 & 0.548 && 0.036 & 0.411 &  6.93$\times 10^{54}$\\ 
S230522n${}^\star$	& BBH	& 24225	& 2928 $\pm$ 1080 &  & 0.995 & 0.471 && 0.199 & 0.228 &  4.64$\times 10^{54}$\\ 
S230529ay   & NSBH	& 24533	& 197 $\pm$ 62 &  & 0.377 & 0.441 && 0.080 & 0.340 &  3.72$\times 10^{52}$\\ 
S230601bf	& BBH	& 2531	& 3565 $\pm$ 1260 &  & 0.819 & 0.193 && 0.536 & 0.088 &  2.15$\times 10^{54}$\\ 
S230605o	& BBH	& 1077	& 1067 $\pm$ 333 &  & 0.786 & 0.094 && 1.000 & 0.094 &  2.21$\times 10^{53}$\\ 
S230606d	& BBH	& 1221	& 2545 $\pm$ 874 &  & 0.759 & 0.036 && 1.000 & 0.032 &  4.64$\times 10^{53}$\\ 
S230608as	& BBH	& 1694	& 3447 $\pm$ 1079 &  & 0.178 & 0.075 && 0.171 & 0.048 &  1.63$\times 10^{54}$\\ 
S230609u	& BBH	& 1287	& 3390 $\pm$ 1125 &  & 0.443 & 0.334 && 1.000 & 0.105 &  3.05$\times 10^{54}$\\ 
S230624av	& BBH	& 1024	& 2124 $\pm$ 682 &  & 0.950 & 0.483 && 1.000 & 0.458 &  4.98$\times 10^{54}$\\ 
S230627c	& NSBH	& 82	& 291 $\pm$ 64 &  & 0.980 & 0.037 && 1.000 & 0.052 &  1.03$\times 10^{52}$\\ 
S230628ax	& BBH	& 705	& 2047 $\pm$ 585 &  & 0.985 & 0.039 && 1.000 & 0.042 &  4.64$\times 10^{53}$\\ 
S230630am	& BBH	& 3965	& 5336 $\pm$ 2001 &  & 0.572 & 0.197 && 0.176 & 0.235 &  2.0$\times 10^{55}$\\ 
S230630bq	& BBH	& 1215	& 999 $\pm$ 286 &  & 0.370 & 0.064 && 0.258 & 0.063 &  1.42$\times 10^{53}$\\ 
S230702an	& BBH	& 2267	& 2428 $\pm$ 849 &  & 0.976 & 0.155 && 1.000 & 0.058 &  6.58$\times 10^{53}$\\ 
S230704f	& BBH	& 1700	& 2759 $\pm$ 992 &  & 0.396 & 0.274 && 0.020 & 0.378 &  6.93$\times 10^{54}$\\ 
S230706ah	& BBH	& 1497	& 1962 $\pm$ 594 &  & 0.580 & 0.277 && 0.349 & 0.214 &  1.75$\times 10^{54}$\\ 
S230707ai	& BBH	& 3181	& 4074 $\pm$ 1485 &  & 0.996 & 0.150 && 1.000 & 0.099 &  4.0$\times 10^{54}$\\ 
S230708cf	& BBH	& 2032	& 3336 $\pm$ 1076 &  & 0.851 & 0.467 && 1.000 & 0.441 &  1.57$\times 10^{55}$\\ 
S230708t	& BBH	& 1227	& 3010 $\pm$ 988 &  & 0.326 & 0.058 && 0.099 & 0.058 &  1.42$\times 10^{54}$\\ 
S230708z	& BBH	& 3373	& 4647 $\pm$ 1696 &  & 0.412 & 0.666 && 0.294 & 0.235 &  1.42$\times 10^{55}$\\ 
S230709bi	& BBH	& 2644	& 4364 $\pm$ 1585 &  & 0.990 & 0.161 && 1.000 & 0.099 &  3.73$\times 10^{54}$\\ 
S230723ac	& BBH	& 1117	& 1551 $\pm$ 436 &  & 0.979 & 0.486 && 1.000 & 0.456 &  2.85$\times 10^{54}$\\ 
S230726a	& BBH	& 27773	& 2132 $\pm$ 714 &  & 0.996 & 0.278 && 0.704 & 0.032 &  2.84$\times 10^{53}$\\ 
S230729z	& BBH	& 1945	& 1495 $\pm$ 444 &  & 0.452 & 0.300 && 0.349 & 0.274 &  1.23$\times 10^{54}$\\ 
S230731an	& BBH	& 599	& 1001 $\pm$ 242 &  & 0.013 & 0.842 && 0.051 & 0.373 &  9.45$\times 10^{53}$\\ 
S230802aq	& BBH	& 25884	& 576 $\pm$ 246 &  & 0.977 & 0.383 && 0.441 & 0.269 &  2.56$\times 10^{53}$\\ 
S230805x	& BBH	& 2094	& 3305 $\pm$ 1113 &  & 0.160 & 0.696 && 0.227 & 0.198 &  5.32$\times 10^{54}$\\ 
S230806ak	& BBH	& 3715	& 5423 $\pm$ 1862 &  & 0.992 & 0.428 && 1.000 & 0.378 &  2.81$\times 10^{55}$\\ 
S230807f	& BBH	& 5436	& 5272 $\pm$ 1900 &  & 0.821 & 0.505 && 1.000 & 0.407 &  2.91$\times 10^{55}$\\ 
S230811n	& BBH	& 810	& 1905 $\pm$ 672 &  & 0.560 & 0.293 && 0.126 & 0.315 &  3.27$\times 10^{54}$\\ 
S230814ah	& BBH	& 25259	& 330 $\pm$ 105 &  & 0.958 & 0.278 && 0.700 & 0.070 &  1.9$\times 10^{52}$\\ 
S230814r	& BBH	& 3389	& 3788 $\pm$ 1416 &  & 0.377 & 0.069 && 0.117 & 0.073 &  2.48$\times 10^{54}$\\ 
S230819ax	& BBH	& 4044	& 4216 $\pm$ 1645 &  & 0.376 & 0.161 && 0.282 & 0.074 &  3.05$\times 10^{54}$\\ 
S230820bq	& BBH	& 1373	& 3600 $\pm$ 1437 &  & 0.980 & 0.449 && 1.000 & 0.470 &  1.71$\times 10^{55}$\\ 
S230822bm	& BBH	& 3974	& 5154 $\pm$ 1771 &  & 0.823 & 0.076 && 0.834 & 0.053 &  3.51$\times 10^{54}$\\ 
S230824r	& BBH	& 3279	& 4701 $\pm$ 1563 &  & 0.987 & 0.203 && 1.000 & 0.169 &  8.3$\times 10^{54}$\\ 
S230825k	& BBH	& 3012	& 5283 $\pm$ 2117 &  & 0.261 & 1.040 && 0.223 & 0.301 &  1.87$\times 10^{55}$\\ 
S230831e	& BBH	& 3803	& 4900 $\pm$ 2126 &  & 0.402 & 0.528 && 0.243 & 0.200 &  1.07$\times 10^{55}$\\ 
S230904n	& BBH	& 2015	& 1095 $\pm$ 327 &  & 0.008 & 0.367 && 0.256 & 0.064 &  2.33$\times 10^{53}$\\ 
S230911ae	& BBH	& 27758	& 1623 $\pm$ 584 &  & 0.996 & 0.365 && 0.221 & 0.241 &  1.32$\times 10^{54}$\\ 
S230914ak	& BBH	& 1532	& 2676 $\pm$ 827 &  & 0.092 & 0.607 && 0.063 & 0.292 &  4.98$\times 10^{54}$\\ 
S230919bj	& BBH	& 708	& 1491 $\pm$ 402 &  & 0.357 & 0.256 && 0.276 & 0.217 &  1.52$\times 10^{54}$\\ 
S230920al	& BBH	& 2180	& 3139 $\pm$ 1003 &  & 0.706 & 0.094 && 0.354 & 0.111 &  2.85$\times 10^{54}$\\ 
S230922g	& BBH	& 324	& 1491 $\pm$ 443 &  & 0.628 & 0.281 && 1.000 & 0.274 &  1.87$\times 10^{54}$\\ 
S230922q	& BBH	& 4658	& 6653 $\pm$ 2348 &  & 0.714 & 0.210 && 0.023 & 0.447 &  3.41$\times 10^{55}$\\ 
S230924an	& BBH	& 835	& 2358 $\pm$ 596 &  & 0.805 & 0.042 && 0.289 & 0.050 &  5.72$\times 10^{53}$\\ 
S230927be	& BBH	& 298	& 1059 $\pm$ 289 &  & 0.996 & 0.128 && 1.000 & 0.206 &  7.39$\times 10^{53}$\\ 
S230927l	& BBH	& 1177	& 2966 $\pm$ 1041 &  & 0.979 & 0.490 && 1.000 & 0.429 &  8.9$\times 10^{54}$\\ 
S230928cb	& BBH	& 3102	& 4060 $\pm$ 1553 &  & 0.035 & 0.927 && 0.022 & 0.383 &  1.18$\times 10^{55}$\\ 
S230930al	& BBH	& 3166	& 4902 $\pm$ 1671 &  & 0.291 & 0.800 && 0.201 & 0.249 &  1.56$\times 10^{55}$\\ 
S231001aq	& BBH	& 3181	& 4425 $\pm$ 1946 &  & 0.521 & 0.606 && 0.144 & 0.601 &  3.76$\times 10^{55}$\\ 
S231005ah	& BBH	& 2497	& 3707 $\pm$ 1335 &  & 0.993 & 0.246 && 1.000 & 0.260 &  9.01$\times 10^{54}$\\ 
S231005j	& BBH	& 5480	& 6417 $\pm$ 2246 &  & 0.665 & 0.239 && 0.212 & 0.240 &  2.72$\times 10^{55}$\\ 
S231008ap	& BBH	& 3102	& 3531 $\pm$ 1320 &  & 0.367 & 0.201 && 0.065 & 0.193 &  5.34$\times 10^{54}$\\ 
S231014r	& BBH	& 1807	& 2857 $\pm$ 903 &  & 0.065 & 0.488 && 0.049 & 0.203 &  4.04$\times 10^{54}$\\ 
S231020ba	& BBH	& 1339	& 1168 $\pm$ 361 &  & 0.993 & 0.044 && 1.000 & 0.049 &  1.8$\times 10^{53}$\\ 
S231020bw	& BBH	& 386	& 2620 $\pm$ 694 &  & 0.969 & 0.023 && 1.000 & 0.029 &  4.98$\times 10^{53}$\\ 
S231028bg	& BBH	& 1207	& 4221 $\pm$ 923 &  & 0.945 & 0.136 && 1.000 & 0.134 &  5.82$\times 10^{54}$\\ 
S231029y	& BBH	& 29972	& 3292 $\pm$ 1313 &  & 0.998 & 0.421 && 0.008 & 0.545 &  8.25$\times 10^{54}$\\ 
S231102w	& BBH	& 2343	& 3493 $\pm$ 1015 &  & 0.554 & 0.243 && 0.445 & 0.159 &  3.76$\times 10^{54}$\\ 
S231104ac	& BBH	& 759	& 1357 $\pm$ 321 &  & 0.997 & 0.196 && 1.000 & 0.222 &  9.33$\times 10^{53}$\\ 
S231108u	& BBH	& 949	& 1986 $\pm$ 494 &  & 0.835 & 0.162 && 1.000 & 0.153 &  1.87$\times 10^{54}$\\ 
S231110g	& BBH	& 636	& 1849 $\pm$ 533 &  & 0.131 & 0.066 && 0.209 & 0.052 &  4.98$\times 10^{53}$\\ 
S231113bb	& BBH	& 2172	& 3260 $\pm$ 1181 &  & 0.459 & 0.037 && 0.539 & 0.029 &  7.05$\times 10^{53}$\\ 
S231113bw	& BBH	& 1713	& 1186 $\pm$ 376 &  & 0.232 & 0.123 && 0.248 & 0.099 &  3.53$\times 10^{53}$\\ 
S231114n	& BBH	& 1267	& 1317 $\pm$ 407 &  & 0.993 & 0.077 && 1.000 & 0.078 &  3.03$\times 10^{53}$\\ 
S231118ab	& BBH	& 2898	& 4353 $\pm$ 1588 &  & 0.735 & 0.091 && 0.708 & 0.067 &  3.51$\times 10^{54}$\\ 
S231118an	& BBH	& 1107	& 1337 $\pm$ 347 &  & 0.187 & 0.474 && 0.073 & 0.303 &  1.32$\times 10^{54}$\\ 
S231118d	& BBH	& 956	& 2110 $\pm$ 585 &  & 0.118 & 0.701 && 0.037 & 0.320 &  3.27$\times 10^{54}$\\ 
S231119u	& BBH	& 5211	& 6597 $\pm$ 2556 &  & 0.464 & 0.580 && 0.008 & 0.565 &  4.94$\times 10^{55}$\\ 
S231123cg	& BBH	& 2714	& 1148 $\pm$ 338 &  & 0.518 & 0.132 && 0.373 & 0.126 &  4.5$\times 10^{53}$\\ 
S231127cg	& BBH	& 3450	& 4425 $\pm$ 1718 &  & 0.775 & 0.095 && 0.708 & 0.088 &  3.51$\times 10^{54}$\\ 
S231129ac	& BBH	& 3089	& 3964 $\pm$ 1513 &  & 0.899 & 0.061 && 1.000 & 0.044 &  1.63$\times 10^{54}$\\ 
S231206ca	& BBH	& 2335	& 3230 $\pm$ 1141 &  & 0.433 & 0.238 && 0.092 & 0.278 &  8.23$\times 10^{54}$\\ 
S231206cc	& BBH	& 342	& 1467 $\pm$ 264 &  & 0.993 & 0.033 && 1.000 & 0.050 &  2.33$\times 10^{53}$\\ 
S231213ap	& BBH	& 1469	& 3861 $\pm$ 1257 &  & 0.990 & 0.037 && 1.000 & 0.043 &  1.5$\times 10^{54}$\\ 
S231223j	& BBH	& 3520	& 4468 $\pm$ 1602 &  & 0.755 & 0.138 && 1.000 & 0.082 &  3.98$\times 10^{54}$\\ 
S231224e	& BBH	& 394	& 863 $\pm$ 213 &  & 0.997 & 0.253 && 1.000 & 0.325 &  6.88$\times 10^{53}$\\ 
S231226av	& BBH	& 199	& 1218 $\pm$ 171 &  & 0.933 & 0.048 && 1.000 & 0.049 &  1.73$\times 10^{53}$\\ 
S231231ag	& BBH	& 27060	& 1066 $\pm$ 339 &  & 0.992 & 0.302 && 0.317 & 0.210 &  5.17$\times 10^{53}$\\ 
S240104bl	& BBH	& 27948	& 1978 $\pm$ 618 &  & 0.992 & 0.557 && 0.376 & 0.134 &  1.45$\times 10^{54}$\\ 
S240107b	& BBH	& 4143	& 6089 $\pm$ 2429 &  & 0.684 & 0.533 && 0.247 & 0.500 &  5.7$\times 10^{55}$\\ 
S240109a	& BBH	& 28048	& 1594 $\pm$ 567 &  & 0.992 & 0.441 && 0.278 & 0.143 &  6.39$\times 10^{53}$\\  
\enddata
\raggedright
\vspace{2pt}
\tablecomments{\,${}^\star$The events S230522a and S230522n were one detector events from ER15, and were only published as Preliminary maps (did not receive advOK status). }

\label{tab:1000s_results}
\end{deluxetable*}

\clearpage
\section{Skymap gallery} \label{app:skymaps}
\begin{figure}[b]
    \centering
    \includegraphics[width=0.32\linewidth]{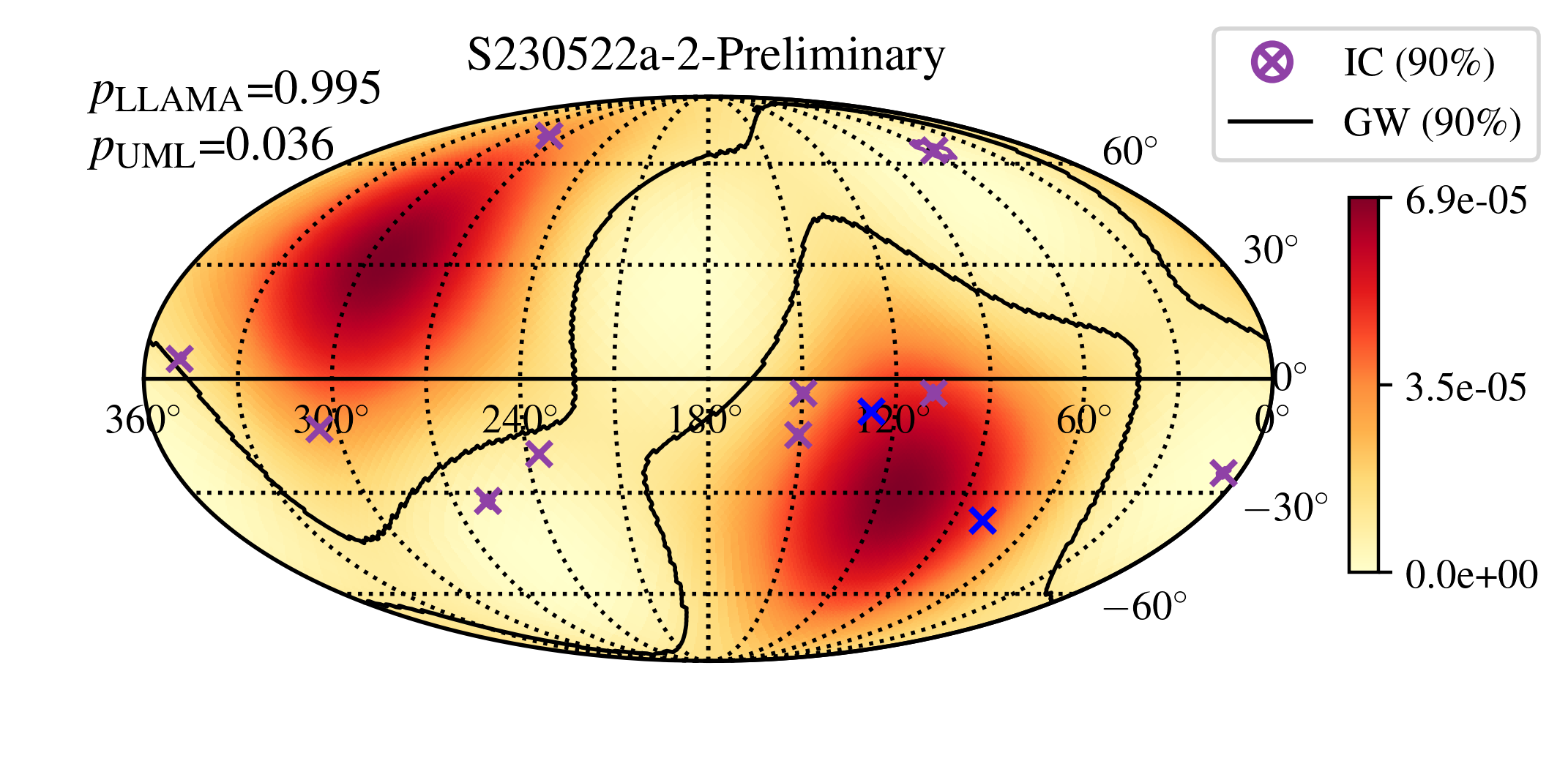}
    \includegraphics[width=0.32\linewidth]{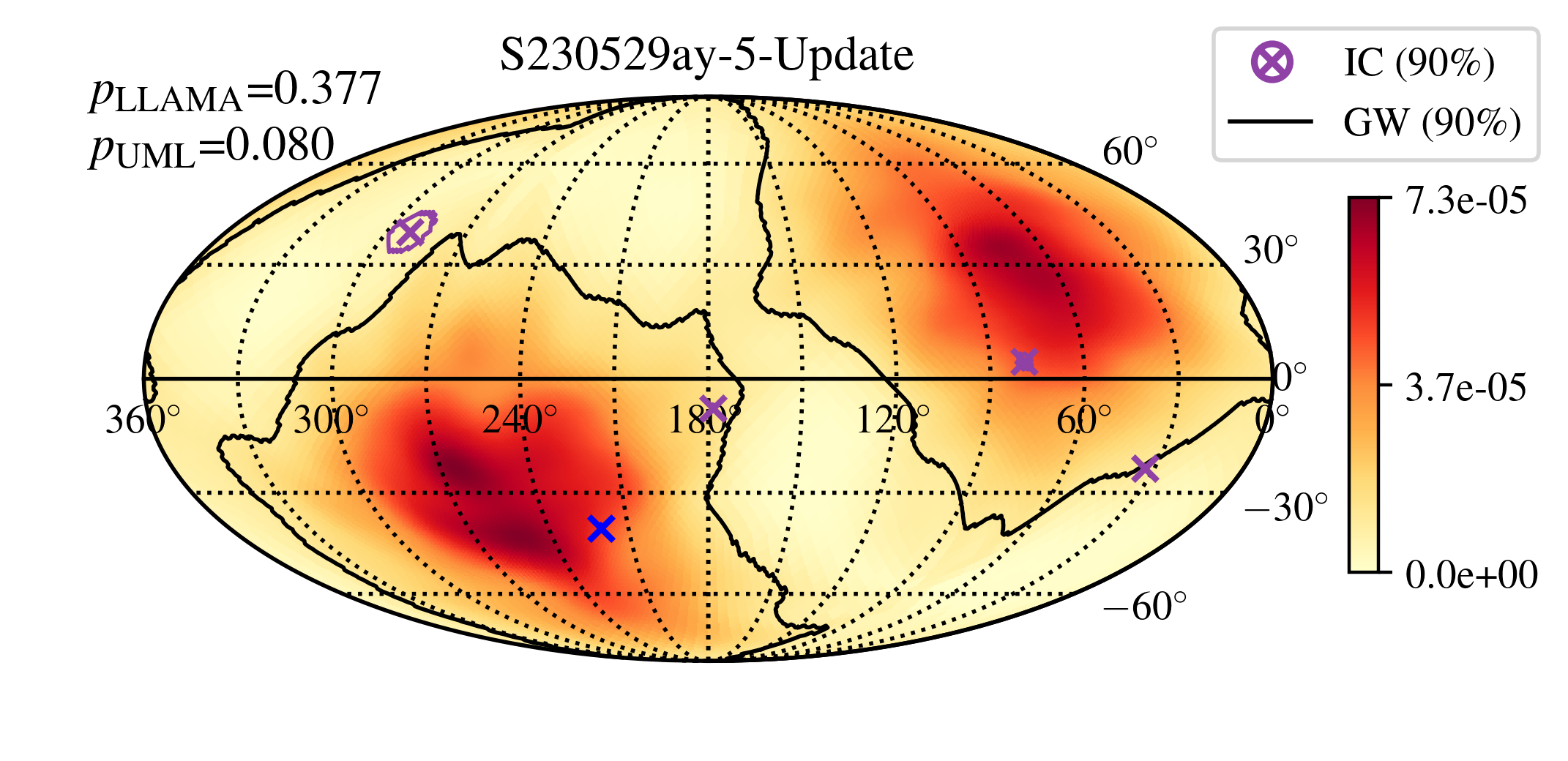}
    \includegraphics[width=0.32\linewidth]{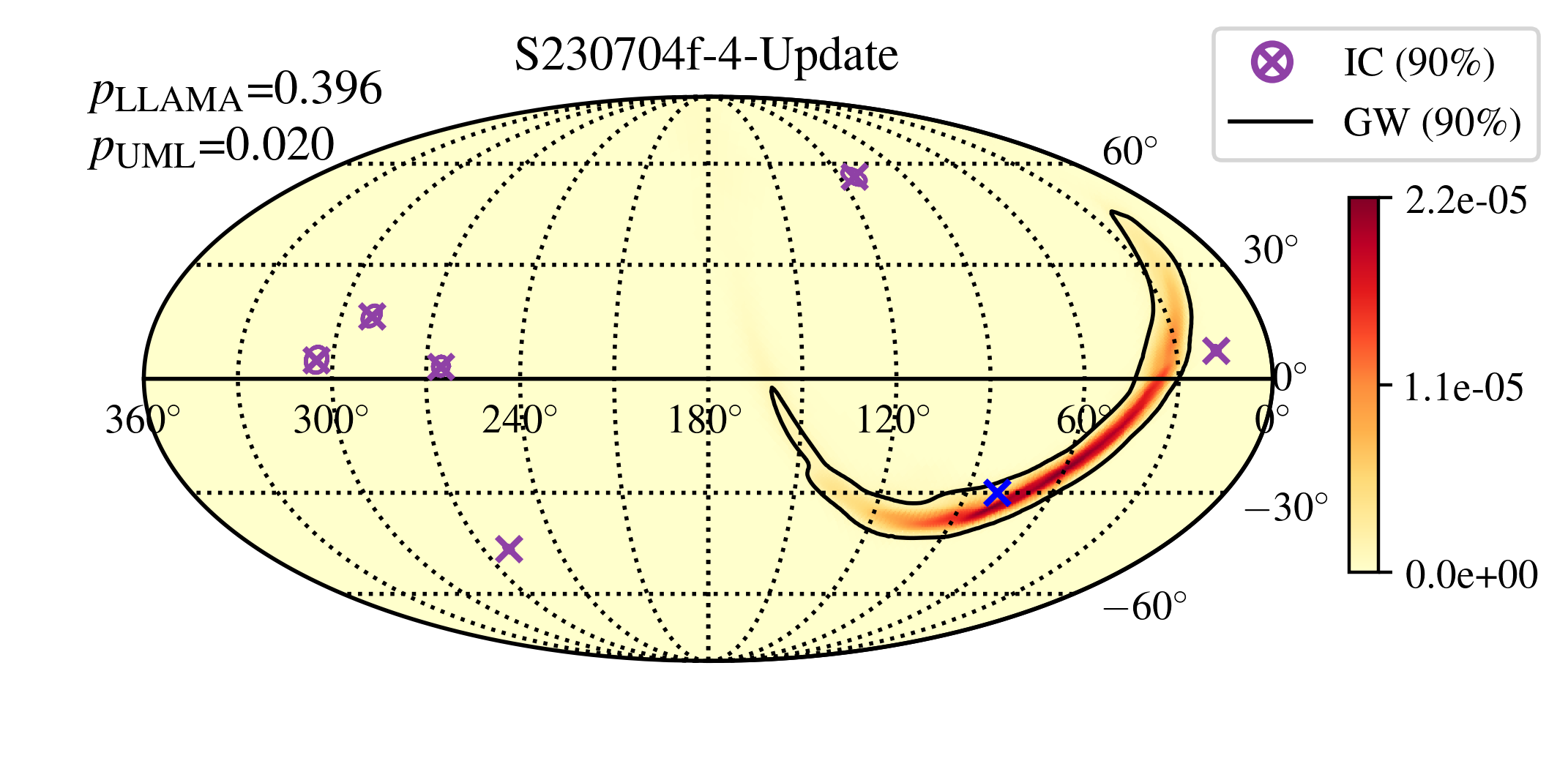}

    \includegraphics[width=0.32\linewidth]{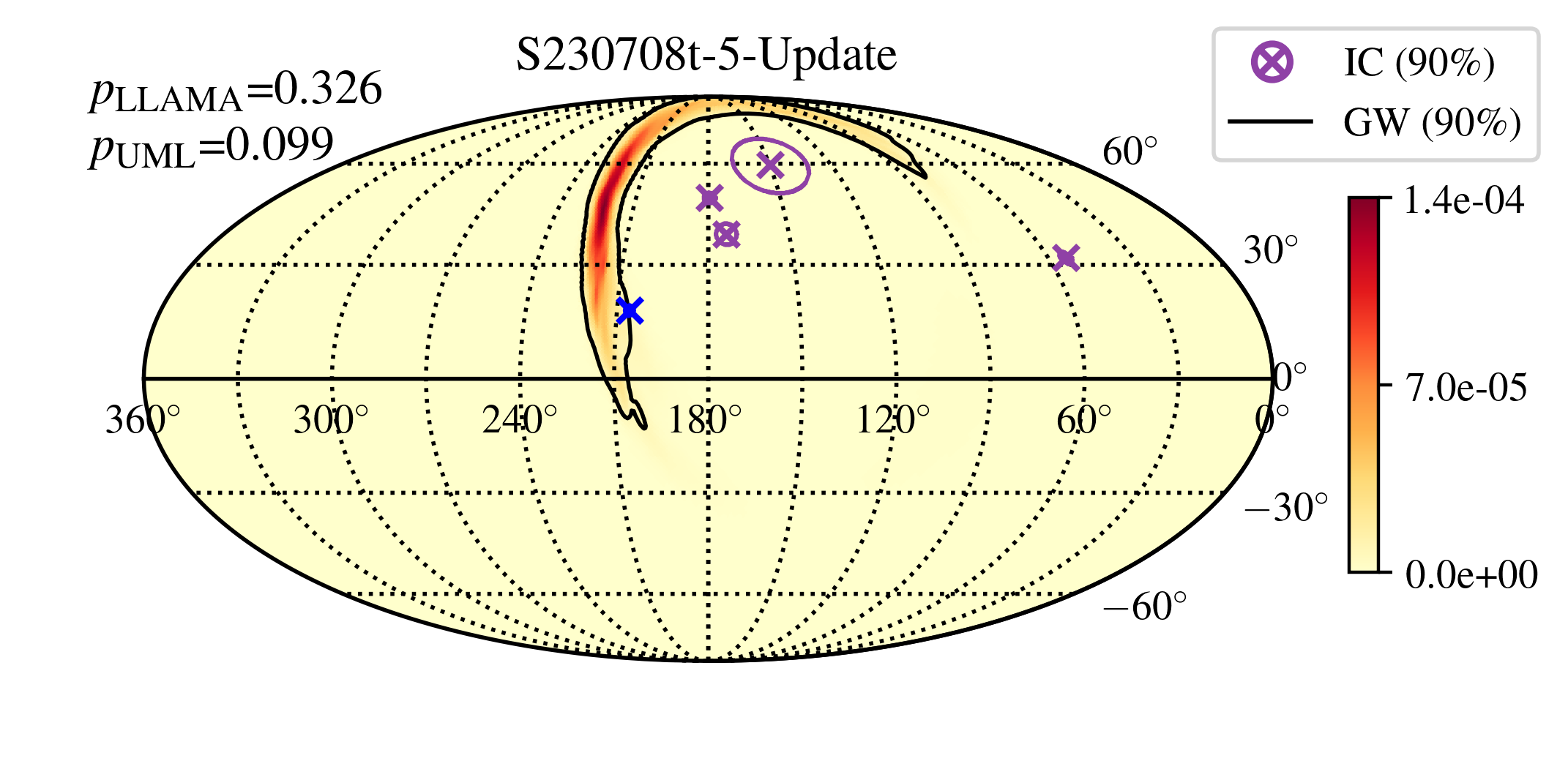}
    \includegraphics[width=0.32\linewidth]{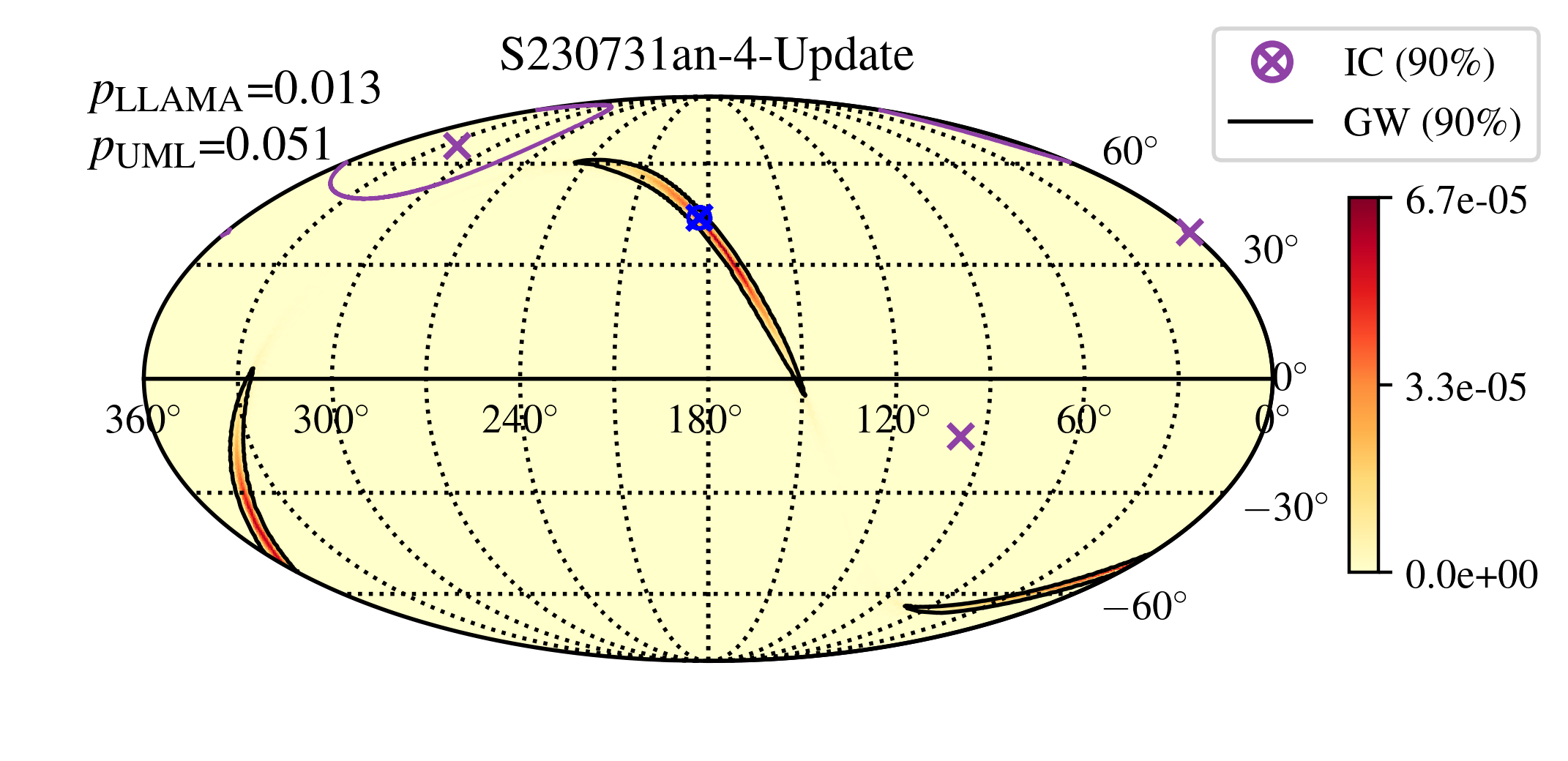}
    \includegraphics[width=0.32\linewidth]{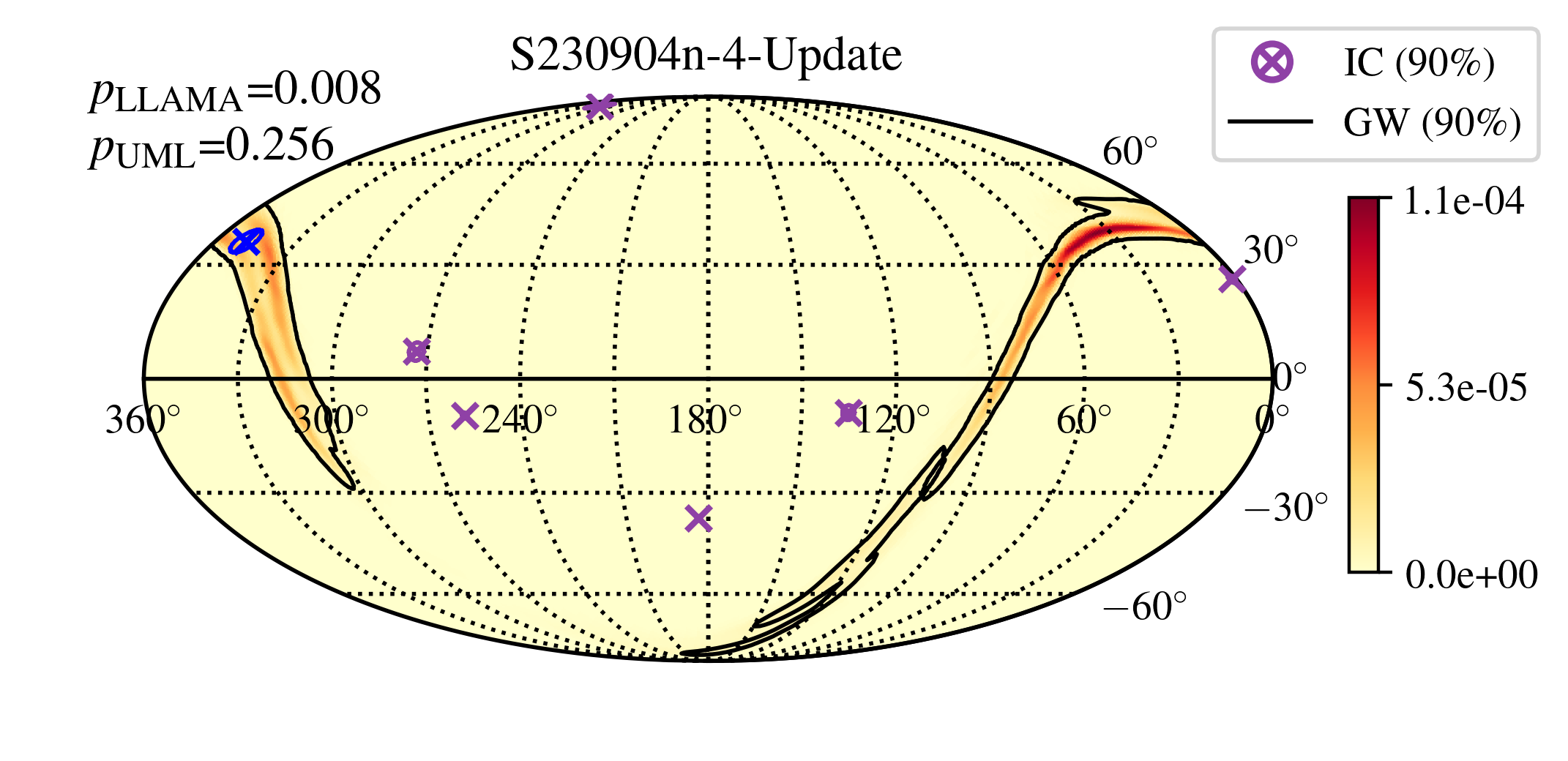}

    \includegraphics[width=0.32\linewidth]{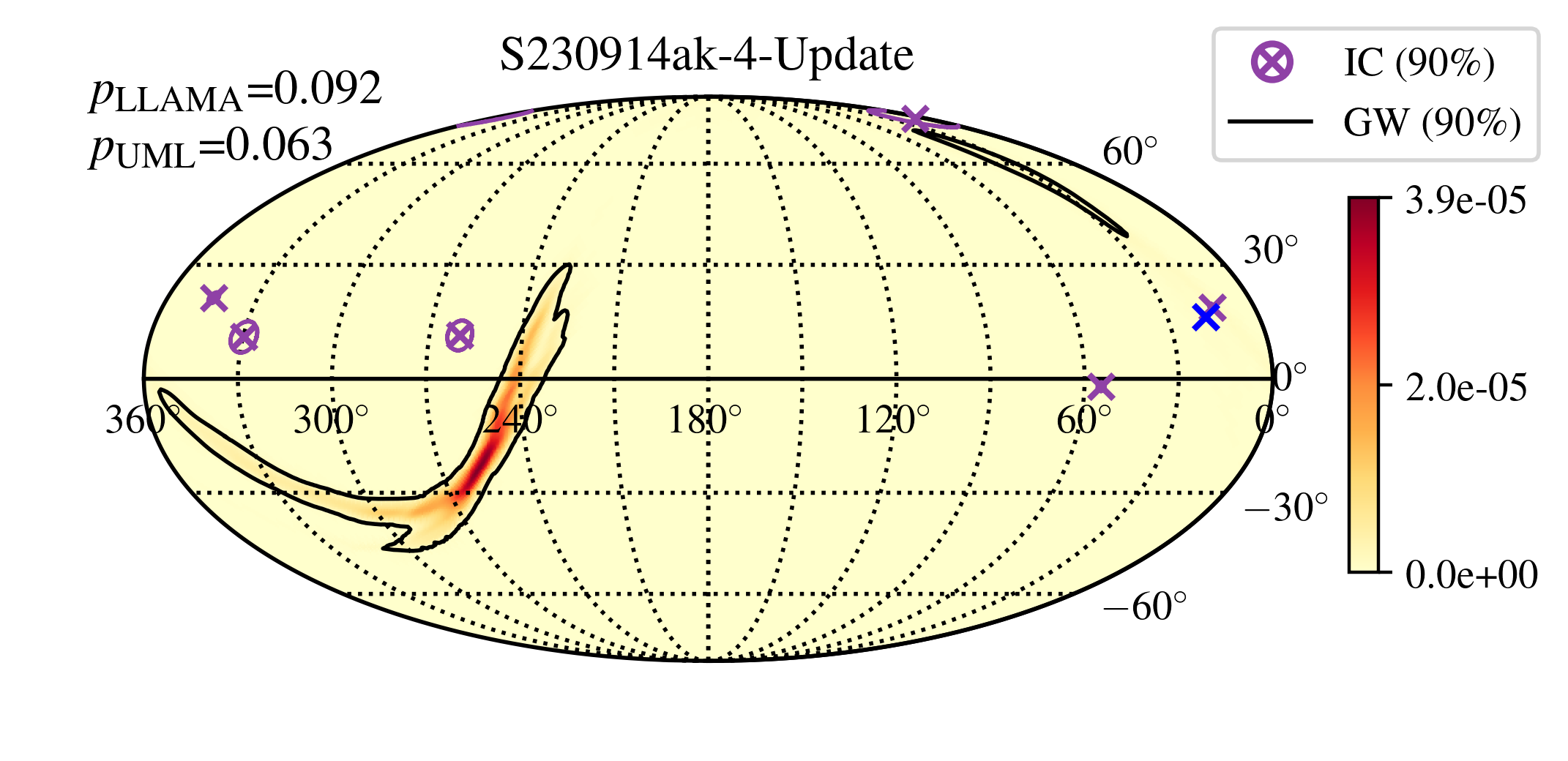}
    \includegraphics[width=0.32\linewidth]{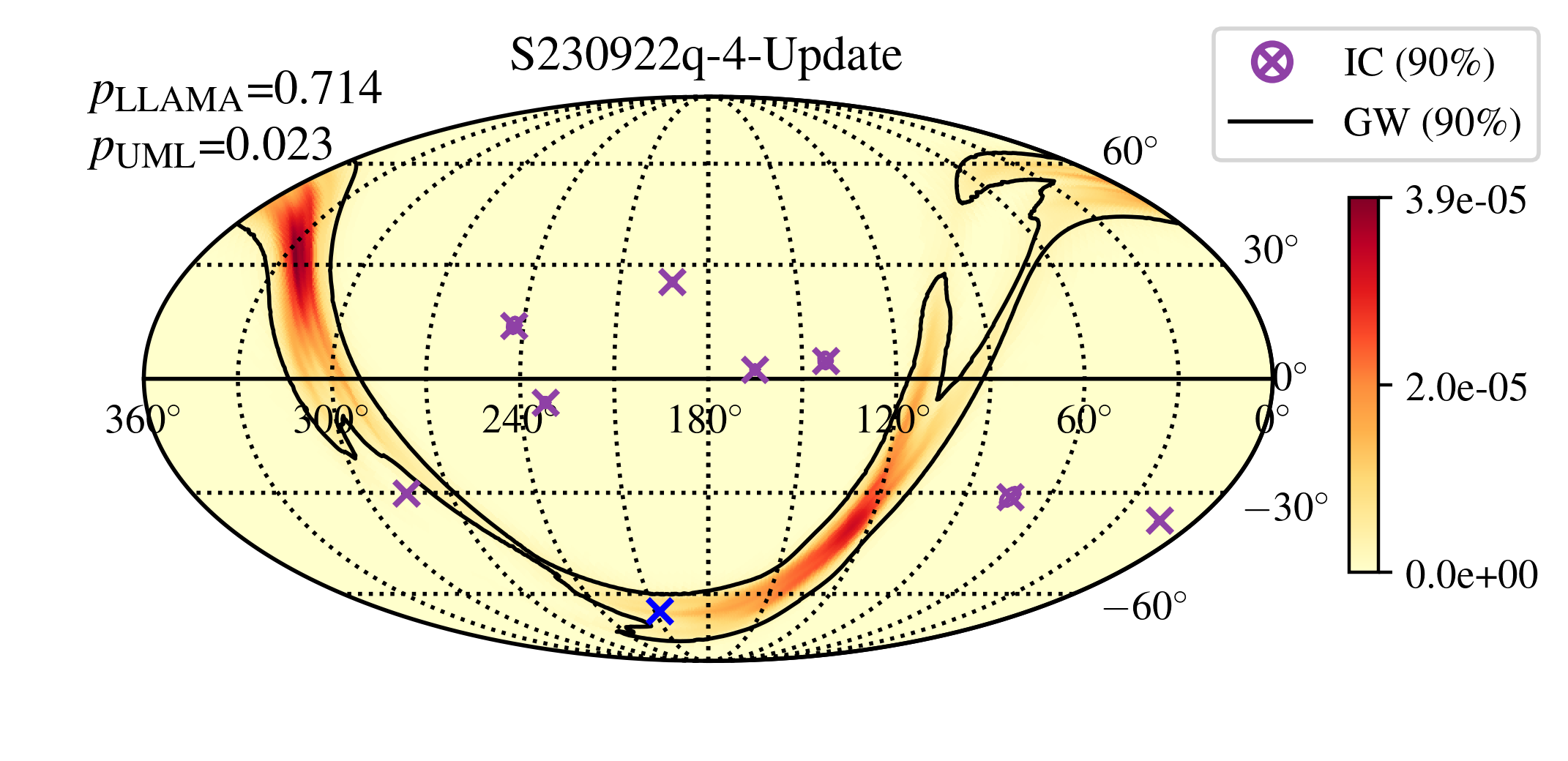}
    \includegraphics[width=0.32\linewidth]{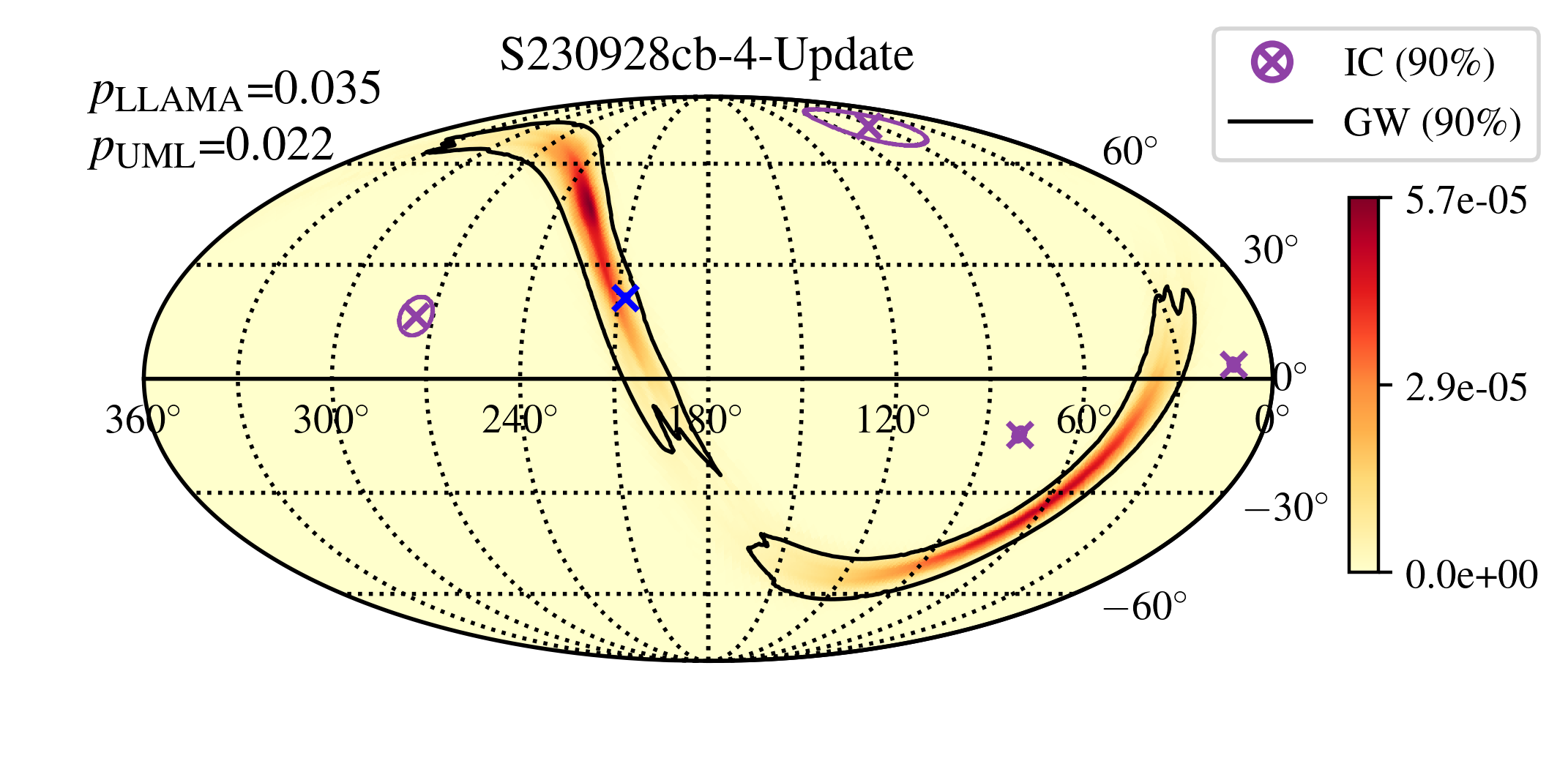}

    \includegraphics[width=0.32\linewidth]{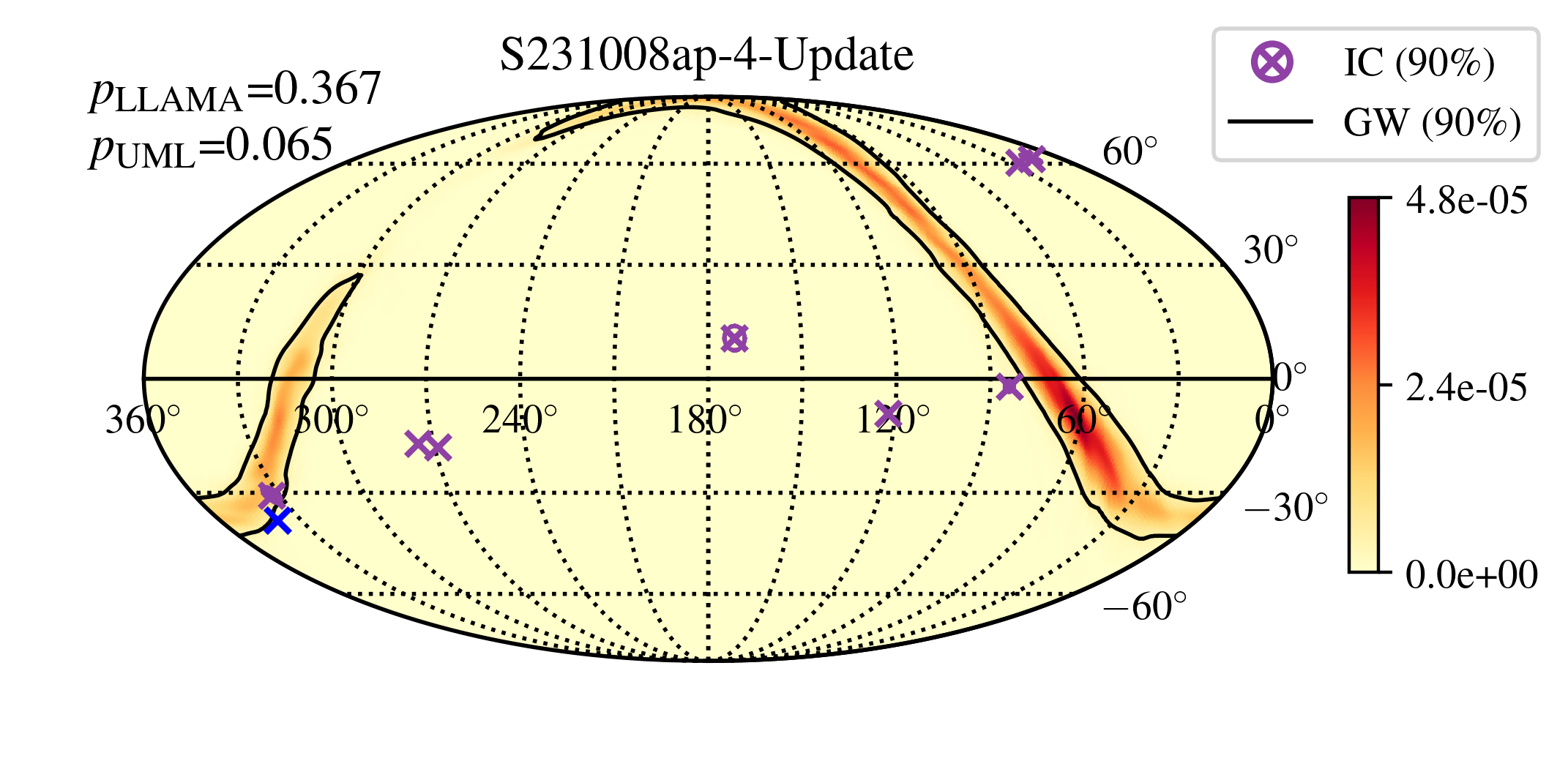}
    \includegraphics[width=0.32\linewidth]{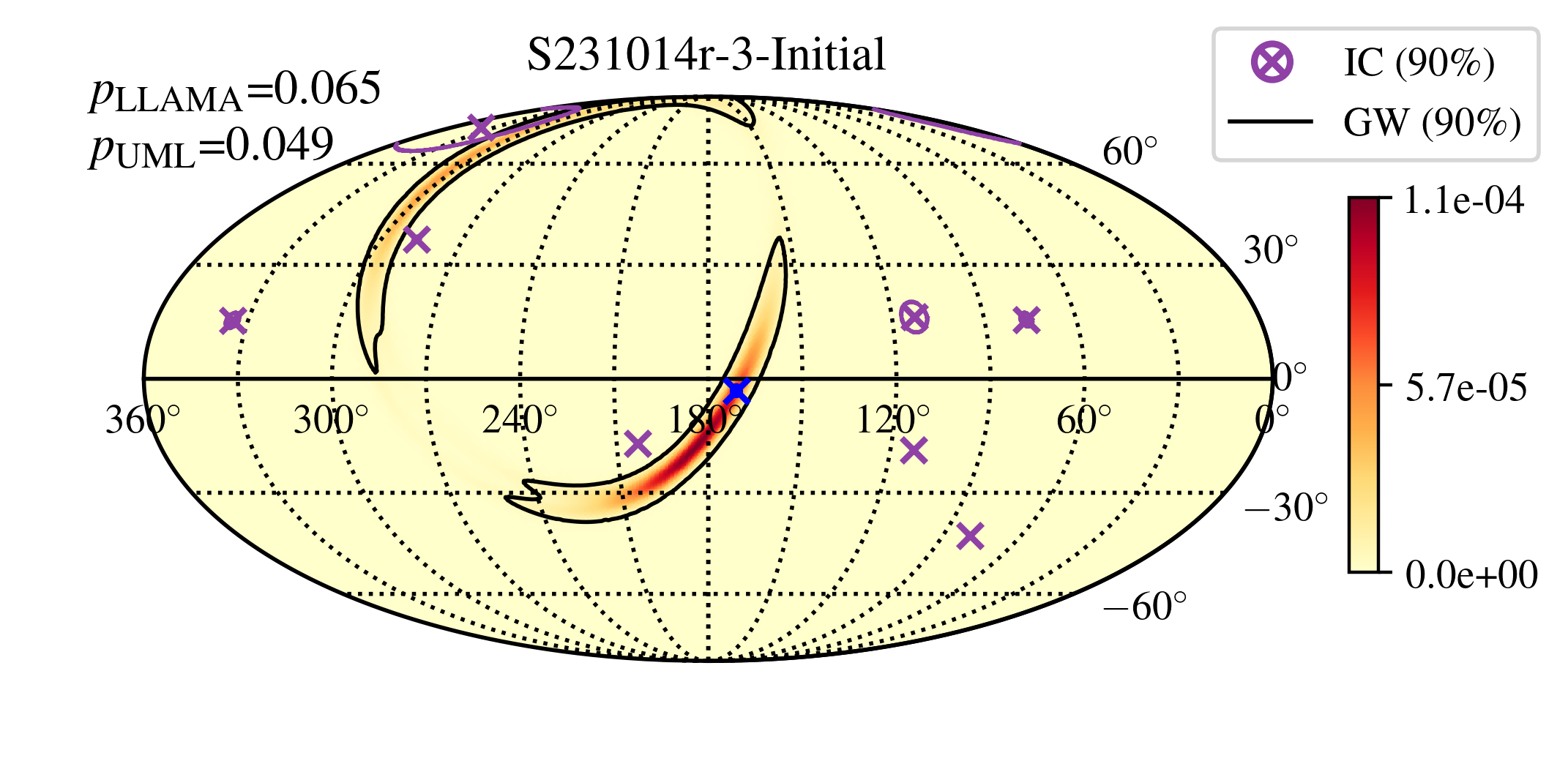}
    \includegraphics[width=0.32\linewidth]{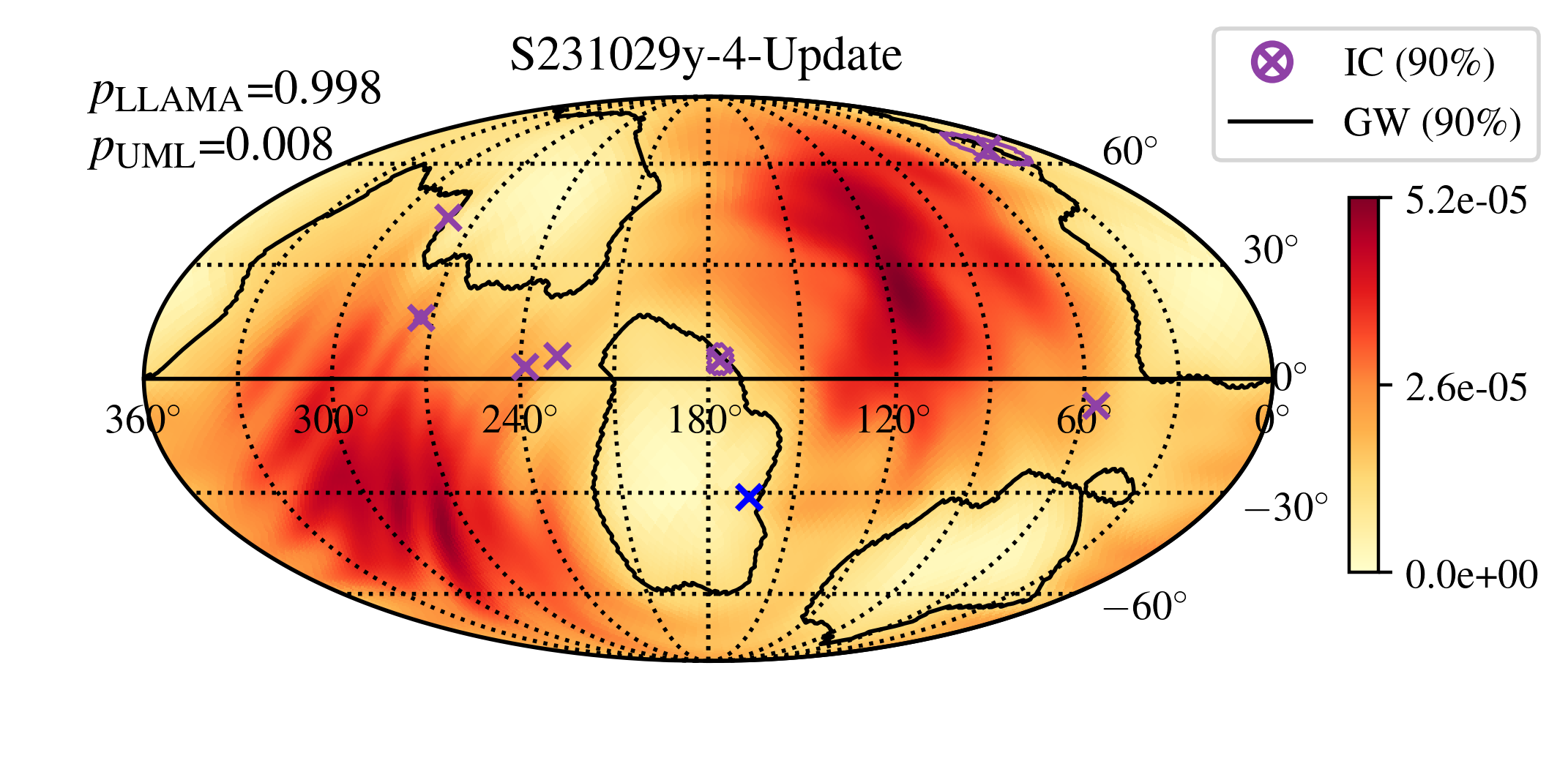}

    \includegraphics[width=0.32\linewidth]{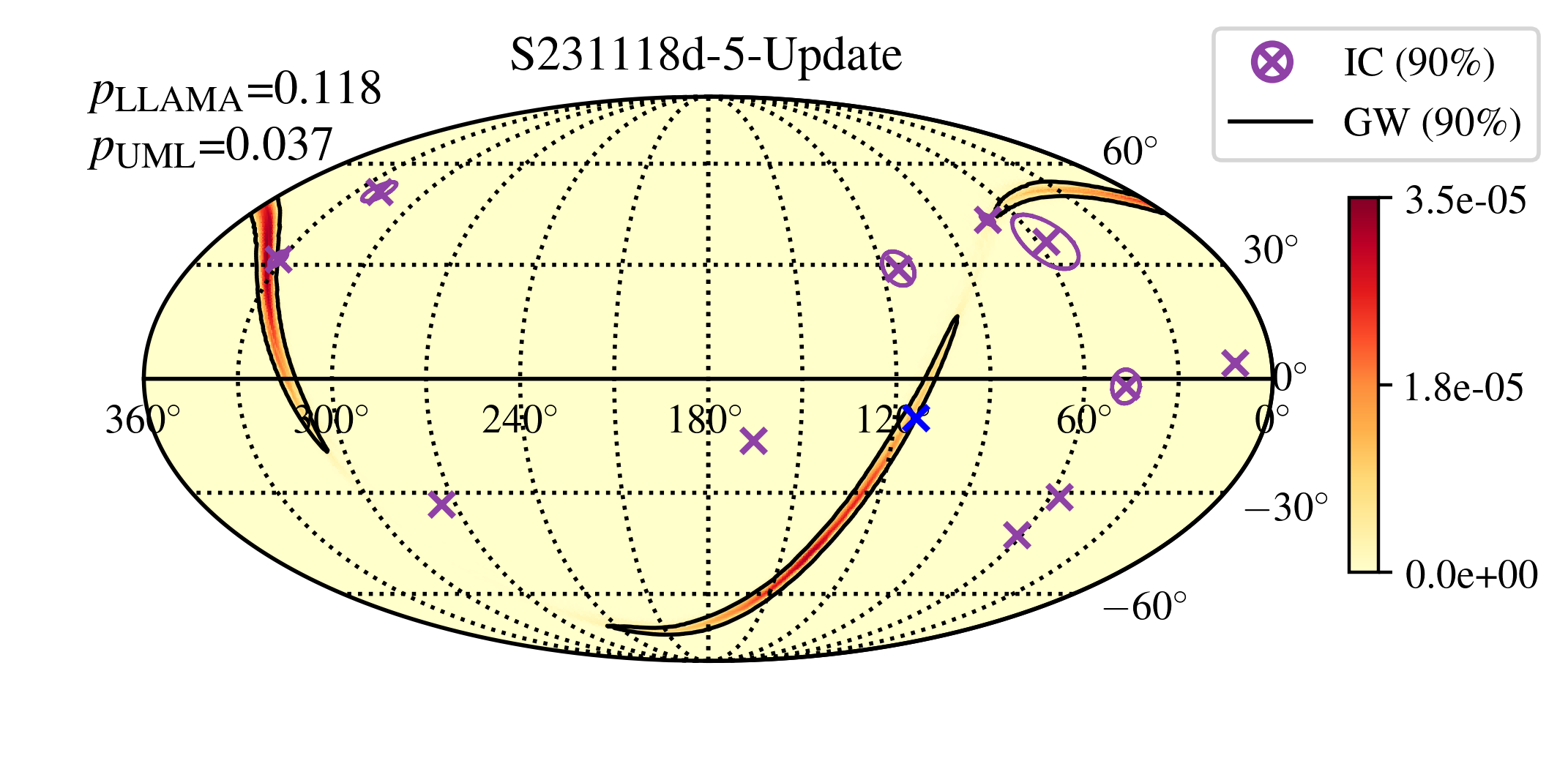}
    \includegraphics[width=0.32\linewidth]{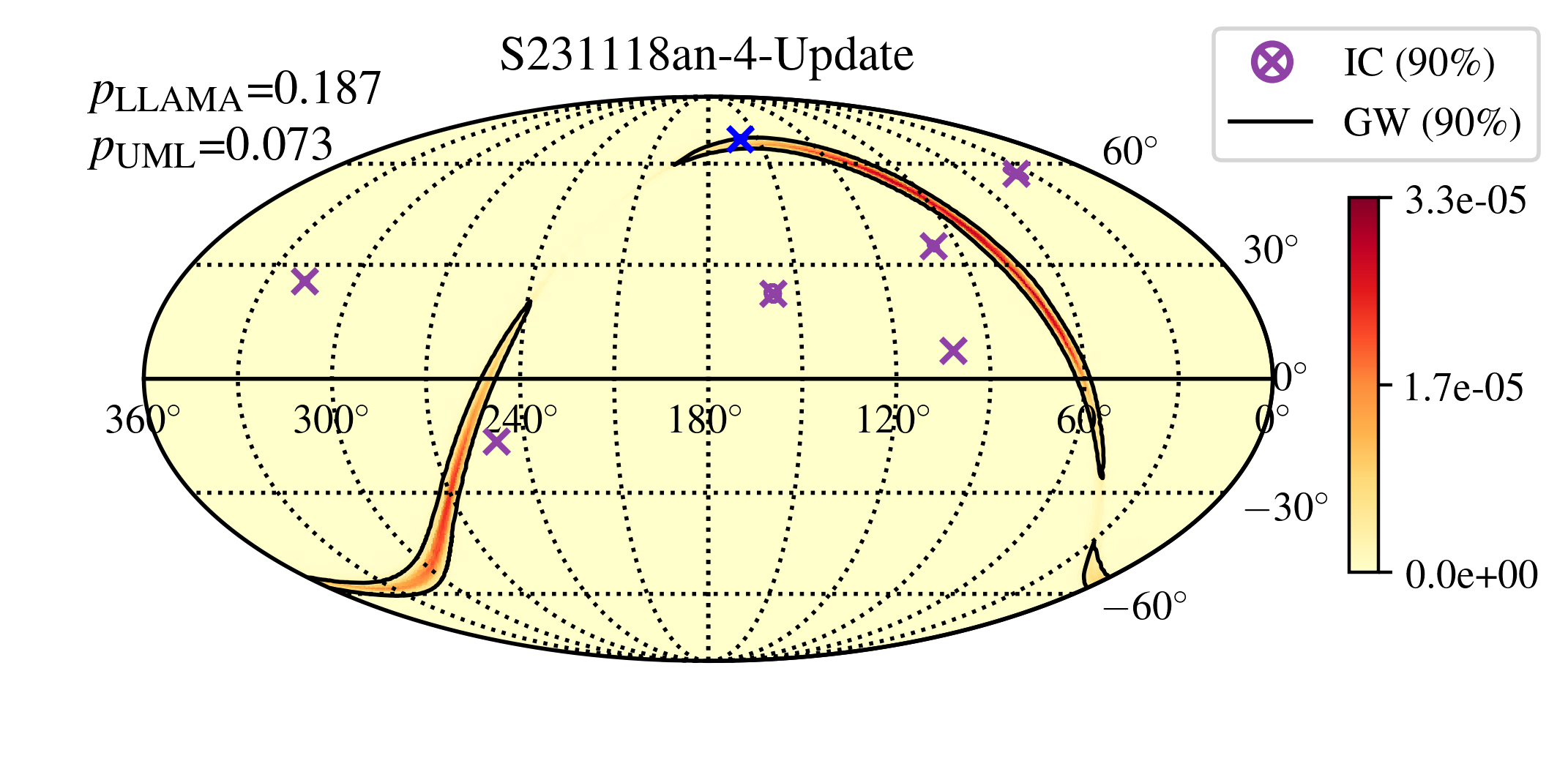}
    \includegraphics[width=0.32\linewidth]{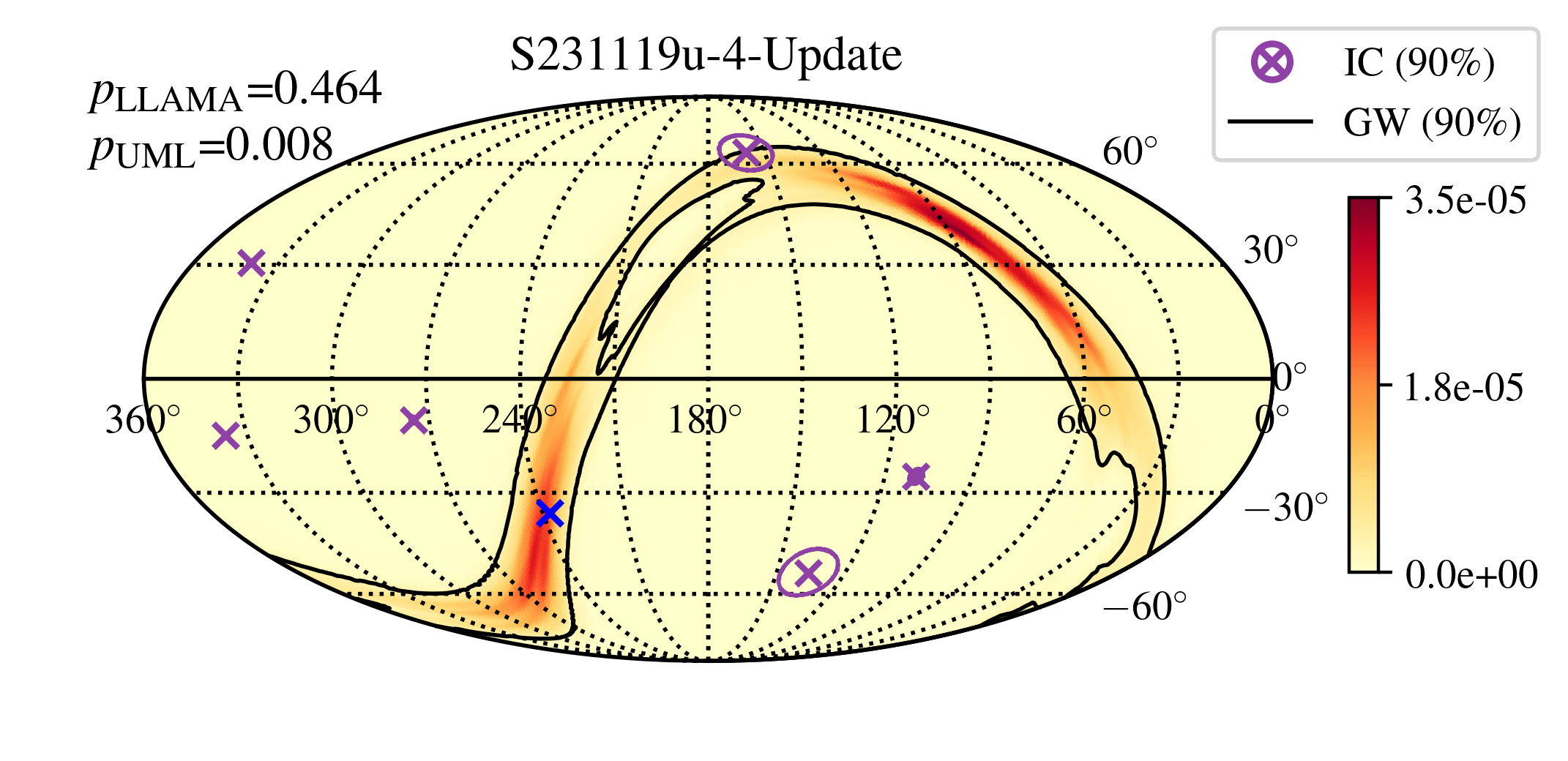}
    
    \includegraphics[width=0.32\linewidth]{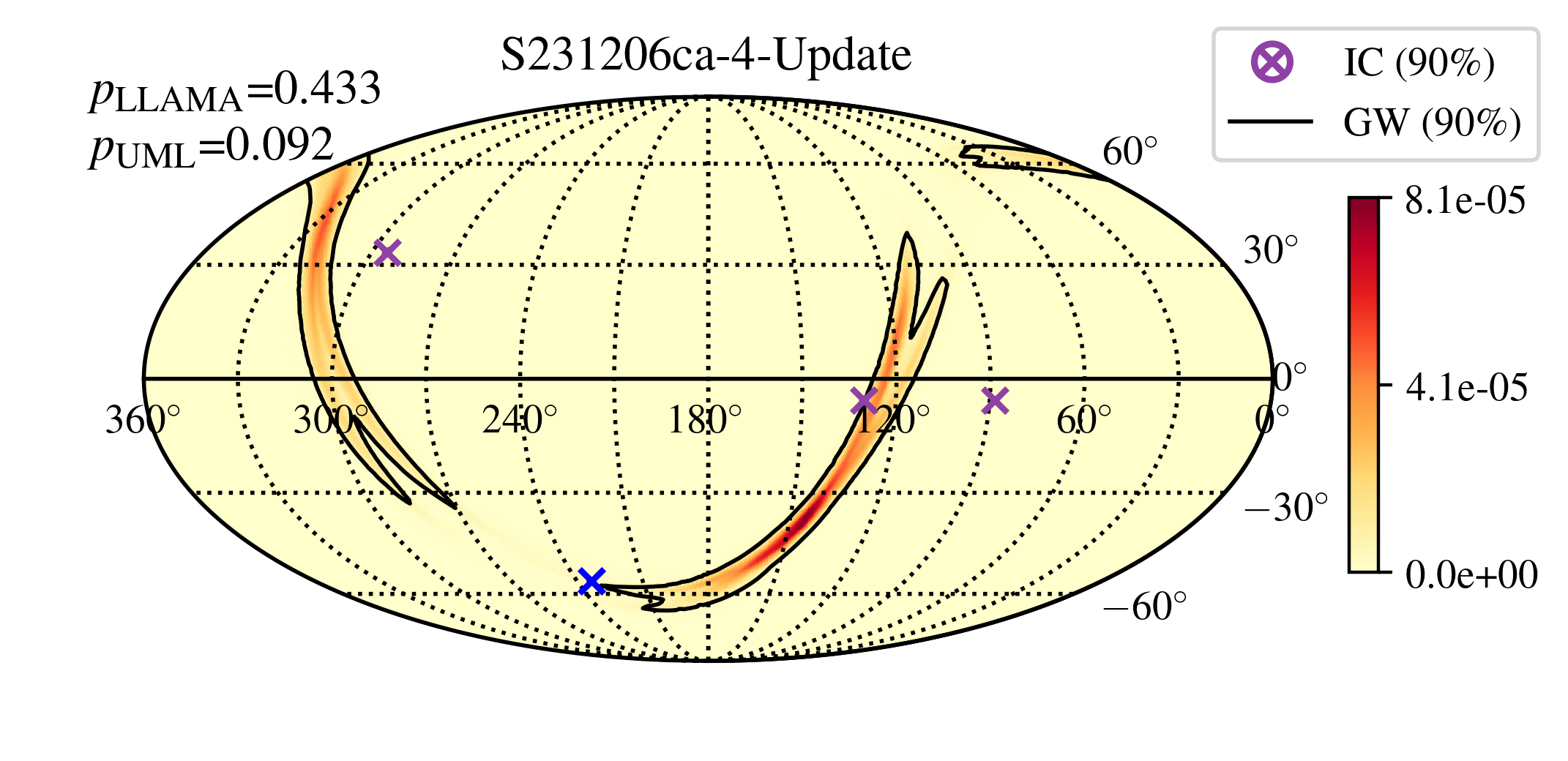}
    
    \caption{Individual skymaps for the $\pm500$ second timescale follow-up. Probability localization regions for the GW events are shown in the colorscale, with the 90\% containment contour shown in black. Overlaid are the IceCube neutrino events, with blue crosses representing the best-fit neutrino direction and the circularized 90\% containment error region in blue circles. All significant alerts which had an overall $p$-value in either analysis less than 0.1 in either analysis are shown, with $p$-values for each alert in the two pipelines annotated on each panel. Individual neutrino events which had a per-event $p$-value less than 10\% are indicated in blue, while other neutrinos in the on-time window are indicated in darker purple. }
    \label{fig:skymap-gallery}
\end{figure}

We plot all of the IceCube events in the $\pm500$ second window analyzed for each event, as well as the GW probability skymap, in Fig.~\ref{fig:skymap-gallery}. Significant events which had $p<0.1$ in either analysis are shown, with the corresponding $p$-values for each event in the upper left corner.

\end{document}